\documentclass[prb,twocolumn,aps,showpacs,superscriptaddress]{revtex4}

\usepackage{graphicx}
\usepackage{dcolumn}
\usepackage{bm}
\usepackage{amsmath}

\begin{document}

\title{Tomonaga-Luttinger liquid correlations and Fabry-Perot interference in conductance and finite-frequency shot noise in a single-walled carbon nanotube}
\author{Patrik Recher}
\altaffiliation{Also at Institute of Industrial Science,
University of Tokyo, 4-6-1 Komaba, Meguro-ku, Tokyo 153-8505,
Japan} \affiliation{Quantum Entanglement Project, E.L. Ginzton
Laboratory, SORST, JST, \\Stanford University, Stanford,
California 94305-4085, USA}

\author{Na Young Kim}
\affiliation{Quantum Entanglement Project, E.L. Ginzton
Laboratory, SORST, JST, \\Stanford University, Stanford,
California 94305-4085, USA}
\author{Yoshihisa Yamamoto}
\altaffiliation{Also at National Institute of Informatics, 2-1-2
Hitotsubashi, Chiyoda-ku, Tokyo 101-8430, Japan}
\affiliation{Quantum Entanglement Project, E.L. Ginzton
Laboratory, SORST, JST, \\Stanford University, Stanford,
California 94305-4085, USA}

\date{April 26, 2006}
\begin{abstract}
We present a detailed theoretical investigation of transport
through a single-walled carbon nanotube (SWNT) in good contact to
metal leads where weak backscattering at the interfaces between
SWNT and source and drain reservoirs gives rise to electronic
Fabry-Perot (FP) oscillations in conductance and shot noise. We
include the electron-electron interaction and the finite length of
the SWNT within the inhomogeneous Tomonaga-Luttinger liquid (TLL)
model and treat the non-equilibrium effects due to an applied bias
voltage within the Keldysh approach.  In low-frequency transport
properties, the TLL effect is apparent mainly via power-law
characteristics as a function of bias voltage or temperature at
energy scales above the finite level spacing of the SWNT. The
FP-frequency is dominated by the non-interacting spin mode
velocity due to two degenerate subbands rather than the
interacting charge velocity. At higher frequencies, the excess
noise is shown to be capable of resolving the splintering of the
transported electrons arising from the mismatch of the
TLL-parameter at the interface between metal reservoirs and SWNT.
This dynamics leads to a periodic shot noise suppression as a
function of frequency and with a period that is determined solely
by the charge velocity. At large bias voltages, these oscillations
are dominant over the ordinary FP-oscillations caused by two weak
backscatterers. This makes shot noise an invaluable tool to
distinguish the two mode velocities in the SWNT.

\end{abstract}

\narrowtext \maketitle

\section{Introduction}
The study of one-dimensional (1D) electronic systems has attracted
much interest due to its unique properties \cite{Giamarchi}. In
1D, electron-electron (e-e) interaction cannot be neglected
anymore but changes the physical properties drastically unlike in
higher dimensional metals which are described successfully by the
Fermi liquid theory.  More specifically, the notion of
quasiparticle excitations completely breaks down in 1D, and the
low-energy excitations are collective charge and spin modes
travelling at different speeds, a phenomenon known as spin-charge
separation.

The low-energy properties of 1D metals have been investigated
successfully within the framework of the Tomonaga-Luttinger liquid
(TLL) theory \cite{Luttinger,Tomonaga}. Recently, a renewed
interest in 1D systems has emerged due to the possibility to
fabricate ideal 1D conductors like carbon nanotubes or
semiconductor quantum wires. Indeed, characteristic predictions of
the TLL-model like power-law renormalized conductance
\cite{Bockrath,West,Yaho} or spin-charge separation \cite{Yacoby}
have been confirmed in the tunneling regime where the 1D system is
well separated from the higher-dimensional reservoirs. Only
recently, transport experiments through single-walled carbon
nanotubes (SWNTs) with an average conductance close to the
theoretical maximum of $G_{0}=4e^2/h$, $h$ is the Planck constant
and $e$ is the electron charge, have been achieved
\cite{HongkunPark, Kong, Kim}. On the theoretical side, Pe\c{c}a
{\it et al.} have calculated the zero temperature conductance for
the model of a SWNT in good contact to two metal reservoirs and
found that the Fabry-Perot (FP) type of interference due to phase
coherent motion within the SWNT is modified by electron-electron
interaction \cite{PBW}. To our knowledge, current noise has not
yet been calculated in this regime of weak backscattering
including the FP-interference between two barriers, see Fig.~1, as
well as two spinfull bands which seems crucial to understand
existing shot noise experiments in SWNT \cite{Kim} where some weak
backscattering at the SWNT-metal reservoir interface cannot be
avoided.

Shot noise is sensitive to temporal correlations of the current
and thus provides additional information about dynamical processes
inside the conductor not accessible in conductance \cite{BuBla}.
In particular, noise is sensitive to the elementary excitations of
the system. In the edge states of the fractional quantum Hall
effect regime, a chiral TLL is realized, where right- and
left-going particles are located at different edges of the sample.
The fractional charge $ge$, with $g$ the TLL-parameter, has been
measured in low-frequency shot noise \cite{fractional} in
agreement with theory \cite{KaneFisher94}. Shot noise measurements
in SWNT are very recent \cite{Roche,Onac, Kim} and no quantitative
analysis of shot noise measurements in the TLL-regime have been
reported so far. In a SWNT right- and left-moving electrons
coexist in the same channel, and consequently electrons can
scatter at the interface between the TLL-system and the
non-interacting reservoirs. Therefore, the physics is expected to
be quite different from its chiral counterpart. One possibility to
model the finite size effect and the influence of the reservoirs
is to use the inhomogeneous TLL-model where the interaction
parameter changes from $g=1$ in the reservoirs to $g<1$ in the
interacting region \cite{SafiSchulz,MaslovStone,Ponomarenko}.
Within this model it has been found that the fractional charge
$ge$ of the TLL cannot be simply extracted from the ratio between
shot noise and backscattered current. It is rather the stable
charge $e$ of the reservoir carriers to which shot noise is
sensitive at low frequencies. This has been concluded for a
single-channel TLL with spin subjected to a random backscattering
potential \cite{Nagaosa} and for a single channel spinless TLL
with a single impurity within the wire
\cite{Trauzettel,Trauzettel1,Dolcini}. We reach here the same
conclusion in the specific FP-setup of Fig.~1. Despite the lack of
a direct measurement of the fractional charge through
low-frequency noise properties, the low-frequency shot noise is
sensitive to interaction since the backscattering off the barriers
is energy dependent leading to power-law dependent noise $S$ and
Fano factor $F=S/eI$, where $I$ denotes the average current.

Recently, it became possible to measure also high-frequency noise
\cite{Deblock,Schoelkopf}. This opens up a way to explore
interaction related effects in an extended parameter range. As
shown in Refs.~22,23,26, the high-frequency noise becomes sensible
to the {\it momentum-conserving} reflections of charge excitations
due to the mismatch of $g$ at the interface between the SWNT and
metal reservoirs which allows to extract further information about
$g$ not contained in low-frequency transport properties. These
multiple reflections are even present without any physical
scatterer \cite{SafiSchulz}, but are only resolved in transport
for frequencies on the order of the interacting level spacing,
i.e. $\hbar\omega\gtrsim \hbar v_{\text{\textsc{f}}} /2Lg$ where
$v_{\text{\textsc{f}}}$ is the Fermi velocity and $L$ is the
length of the interacting region. However, the situation is
different once an impurity is included in the system. Electron
waves can be scattered at the impurity site and interfere with the
transmitted part which is partially backscattered at the interface
due to the inhomogeneity of $g$ which leads also to oscillations
with frequency $v_{\text{\textsc{f}}}/2Lg$ as a function of bias
voltage. This point was noted in several works
\cite{Smitha1,PBW,Dolcini2,Dolcini}. In the experimentally
relevant case of a SWNT with two impurities, this interaction
induced interference is masked by the usual FP-oscillations due to
two scatterers naturally formed at the interface between the SWNT
and metal reservoirs. Since the SWNT has three non-interacting
modes due to spin and subband degeneracy and only one interacting
mode of the total charge carrying the information about $g$, any
oscillation in the bias voltage dependence of conductance or noise
is dominated by the non-interacting spin mode frequency
$v_{\text{\textsc{f}}}/L$. However, as pointed out in Ref.~11,
applying a gate voltage can decrease the amplitude of the ordinary
FP-interference. In that case, small oscillations with frequency
$v_{\text{\textsc{f}}}/2Lg$ remain. The TLL-parameter $g$ is
presumably only weakly dependent on gate voltage
\cite{gatevoltage}. However, in general, applying a gate voltage
can influence $g$ in a TLL due to screening by the gate electrode
\cite{West,Dolcini}. We find now that noise as a function of
frequency $\omega$ and bias voltage $V$ is capable of clearly
discriminating the two oscillation periods of collective modes
present in the SWNT without changing the gate voltage. At high
bias voltages ($eV\gg\hbar\omega, \hbar v_{\text{\textsc{f}}}/L$),
we find that the frequency dependent excess noise shows
oscillations dominated by the charge-mode frequency
$v_{\text{\textsc{f}}}/2Lg$, whereas the bias voltage dependence
exhibits the FP-oscillations dominated by the non-interacting
spin-mode frequency $v_{\text{\textsc{f}}}/L$ whose amplitude is
modulated by $\omega$. This clearly distinguishes the charge
plasmon resonance induced by the finite length $L$ of the
interacting region (SWNT) from the more conventional
FP-interference due to two barriers. The finite frequency noise
therefore could be used to extract both frequency scales which
allows us to extract $g$ without the knowledge of any system
parameters like the position of an impurity
\cite{Trauzettel1,Dolcini} or the fitting to a power-law
\cite{Bockrath, Yaho}. This is highly anticipated since power-laws
can also originate from environmental effects (dynamical Coulomb
blockade) in the same functional form \cite{SafiSaleur}.

The organization of the paper is as follows: in Section II we
introduce the TLL-model of a SWNT with spatially inhomogeneous
TLL-parameter taking into account the effects of the
non-interacting  source and drain electrodes. We then discuss the
inclusion of two weak backscattering potentials situated at the
interfaces between metal electrodes and SWNT. In Section III we
introduce the general framework of a Keldysh functional integral
approach to treat the non-equilibrium effects due to an applied
bias voltage. In Section IV we present the dc conductance to
leading order in the backscattering thereby extending the result
of Ref.~11 to finite temperatures \cite{saficomment}. We also give
some asymptotic analytical results showing the relevant power-laws
and provide numerical results for the generic case. Section V is
devoted to the current noise where we discuss the low-frequency
noise, Fano factor and the general frequency dependence. The
details of the calculations are presented in the appendices.
Sections IV and V close with a discussion of the physical
interpretation of the results. We set $\hbar=1$ in intermediate
steps but restore $\hbar$ in final results.

\section{Model for SWNT coupled to metal reservoirs}
We consider electrons in a SWNT subjected to a repulsive
 Coulomb interaction potential parametrized by $\lambda>0$ with
Hamiltonian density \cite{CBF,Egger}
\begin{align} \label{model1}
{\cal
H}_{\text{\textsc{swnt}}}=&-iv_{\text{\textsc{f}}}\sum\limits_{i=1}^{2}\sum\limits_{s=\uparrow,\downarrow}\left
[\psi_{Ris}^{\dagger}\partial_{x}\psi_{Ris}-\psi_{Lis}^{\dagger}\partial_{x}\psi_{Lis}\right]\nonumber\\
+&\lambda\,\rho_{\rm tot}^{2}(x),
\end{align}
where $\rho_{\rm tot}(x)=
\sum_{i=1}^{2}\sum_{s=\uparrow,\downarrow}(\psi_{Ris}^{\dagger}\psi_{Ris}+\psi_{Lis}^{\dagger}\psi_{Lis})$
is the total charge density and $i=1,2$ denotes the two bands that
cross the Fermi level. The slow-varying parts of the
field-operators for left(L) and right(R) moving electrons can be
expressed in terms of bosonic fields as
\begin{equation}
\label{bd}
\psi_{R/Lis}=\frac{1}{\sqrt{2\pi\Lambda}}\,e^{i(\phi_{is}\pm\theta_{is})},
\end{equation}
which satisfy the commutation relation
$[\phi_{is}(x),\theta_{js'}(x')]=i(\pi/2)\delta_{ij}\delta_{ss'}{\rm
sgn}(x-x')$. This relation implies that
$\Pi_{is}(x)=-(1/\pi)\partial_{x}\phi_{is}(x)$ is the conjugate
momentum to $\theta_{is}(x)$. In Eq.~(\ref{bd}) we have introduced
a short-distance cut-off $\Lambda$ which is on the order of the
lattice spacing \cite{Kleinfactor}. To proceed, it is useful to
define new fields for total charge(spin) and charge(spin)
imbalance between the two bands. We define charge($c$) and
spin($\sigma$) bosonic fields via
$\theta_{ic}=(\theta_{i\uparrow}+\theta_{i\downarrow})/\sqrt{2}$
and
$\theta_{i\sigma}=(\theta_{i\uparrow}-\theta_{i\downarrow})/\sqrt{2}$
and further the symmetric(+) and antisymmetric(--) combinations
$\theta_{\pm\mu}=(\theta_{1\mu}\pm\theta_{2\mu})/\sqrt{2}$,
$\mu=c,\sigma$ and similar for $\phi$-fields. We obtain four
labels \cite{CBF}: $a=\{1=+\rho,2=+\sigma,3=-\rho,4=-\sigma\}$. In
this new basis the Hamiltonian  $H_{\text{\textsc{swnt}}}=\int dx
{\cal H}_{\text{\textsc{swnt}}}$ for the SWNT incorporating the
reservoirs becomes
\begin{align}
\label{bosoH}
H_{\text{\textsc{swnt}}}=&\frac{v_{\text{\textsc{f}}}}{2\pi}\int
dx\,\left[(\partial_{x}\phi_{1})^2+\frac{1}{g^2(x)}(\partial_{x}\theta_{1})^2\right]\nonumber\\
+&\frac{v_{\text{\textsc{f}}}}{2\pi}\sum\limits_{a=2}^{4}\,\int
dx\,\left[(\partial_{x}\phi_{a})^2+(\partial_{x}\theta_{a})^2\right].
\end{align}
The velocity of the collective charge excitations in the SWNT is
$v_{c}=v_{\text{\textsc{f}}}/g$ which is renormalized due to
repulsive e-e interaction in the nanotube. Since the interaction
potential strength $\lambda$ couples only to the total charge
density, only the charge sector $a=1$ is modified by the
TLL-parameter
$g=[v_{\text{\textsc{f}}}/(v_{\text{\textsc{f}}}+(8\lambda/\pi))]^{1/2}$.
We assume $g(x)=g<1$ in the SWNT and $g(x)=1$ in the reservoirs.
The inhomogeneity of $g$ reflects the finite size of the nanotube.
The abrupt change of $g$ at the interfaces between metal
reservoirs and SWNT is considered to be a good approximation to a
smooth transition of $g$ as long as the real length over which $g$
changes is much smaller than the typical wavelengths of the
excitations in the TLL, but larger than the Fermi wavelength or
lattice spacing \cite{Dolcini}. The relation of the bosonic fields
in Eq.~(\ref{bosoH}) to physical quantities can be examined by
looking at products of fermion operators. Using the relation for
normal ordered densities
$n_{R/L}(x)=:\,\psi_{R/Lis}^{\dagger}(x)\psi_{R/Lis}(x)\,:=\pm\,\partial_{x}(\phi_{is}(x)\pm\theta_{is}(x))/2\pi$,
we obtain, e.g.  for the {\it total charge} density, $\rho_{\rm
tot}(x)=(2/\pi)\partial_{x}\theta_{1}(x)$. Of particular interest
is the operator for the charge current. From the continuity
equation we obtain ${\hat I}(x,t)=-e(2/\pi)\dot{\theta}_{1}(x,t)$.

\begin{figure}[h]
\centerline{\includegraphics[width=7.5cm]{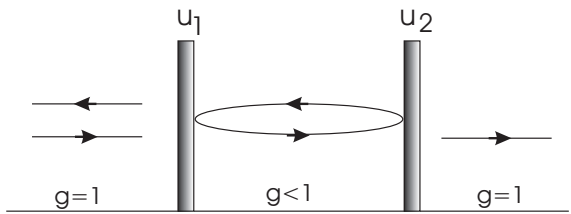}}
\caption{The Fabry-Perot double barrier device: The backscattering
with bare amplitude strengths $u_{1}$ and $u_{2}$ of electrons
primarily takes part at the SWNT-metal reservoir interfaces where
the TLL parameter $g$ changes  from $g$=1 in the leads to $g<1$ in
the nanotube.}
\end{figure}

The backscattering off impurities is assumed to be weak and mainly
happening at the two metal contact-SWNT interfaces which separate
the nanotube from the reservoirs. The form of the backscattering
Hamiltonian is given as
\begin{align}
\label{backh}
H_{\rm bs}=&\sum\limits_{m=1}^{2}\sum_{i,j=1}^{2}\,\sum_{s=\uparrow,\downarrow}\,\tilde{u}_{m}^{ij}\,e^{i(-1)^{m+1}\Delta_{ij}}\nonumber\\
\times&\psi^{\dagger}_{Lis}(x_{m})\psi_{Rjs}(x_{m})+{\rm h.c.}\nonumber\\
=&\sum\limits_{m,i,j=1}^{2}\sum\limits_{s=\pm
1}\,u_{m}^{ij}\,\exp\bigr\{i[\theta_{1m}+s\theta_{2m}\nonumber\\
+&(-1)^{i+1}\delta_{ij}(\theta_{3m}+s\theta_{4m})\nonumber\\
+&(-1)^{i+1}(1-\delta_{ij})(\phi_{3m}+s\phi_{4m})\nonumber\\
+&(-1)^{m+1}\Delta_{ij}]\bigl\}+{\rm h.c.}
\end{align}
In Eq.~(\ref{backh}) we have used
$\theta_{am}\equiv\theta_{a}(x_{m})$ and similarly for $\phi_{am}$
with $x_{1,2}=\mp L/2$ denoting the positions of the two barriers.
We have further defined
$u^{ij}_{m}=\tilde{u}_{m}^{ij}/2\pi\Lambda$ which are real valued
and have the dimension of energy. The backscattering phase for the
scattering of  a right-moving electron with band-index $j$ to a
left-moving electron with band-index $i$
 is denoted by  $(-1)^{m+1}\Delta_{ij}$. Its dependence on the contact label $m=1,2$ reflects the mirror symmetry of the two SWNT-metal reservoir  interfaces with respect to
 $x=0$. We further assume that these phases are energy independent.

Next, we discuss the inclusion of a gate voltage $V_{\rm g}$ which
gives rise to a Hamiltonian density proportional to the total
charge density $H'\propto \rho_{\rm tot}\,V_{\rm
g}=(2/\pi)(\partial_{x}\theta_{1})V_{\rm g}$. This linear term in
the Hamiltonian can be eliminated by performing the linear shift
\cite{PBW} $\theta_{1}\rightarrow\theta_{1}-V_{\rm g}x$ which
leaves the quadratic Hamiltonian Eq.~(\ref{bosoH}) unchanged (up
to an irrelevant constant) but changes $H_{\rm bs}$ where
$\theta_{1}$ is replaced by $\theta_{1}-V_{\rm g}x$. Note that
applying a gate voltage induces a shift of the Fermi level in the
SWNT and metal contacts.
\section{The transport theory}
In this section we derive the general framework for calculating
the current and current noise in non-equilibrium within the
Keldysh functional approach. We start with the system Hamiltonian
$H=H_{\text{\textsc{swnt}}}+H_{\rm bs}$ and treat $H_{\rm bs}$ as
the perturbation. The average of an observable ${\cal O}$ is
$\langle\hat{{\cal O}}(t)\rangle={\rm Tr}[{\hat \rho} \hat{{\cal
O}}(t)]$ where $\hat{{\cal O}}(t)=e^{iH(t-t_{0})}\hat{{\cal
O}}e^{-iH(t-t_{0})}$, ${\hat \rho}$ is the density matrix at time
$t_{0}$ before $H_{\rm bs}$ is switched on and Tr means trace.
\begin{figure}[b]
\label{Kcontour}
\centerline{\includegraphics[width=7.5cm]{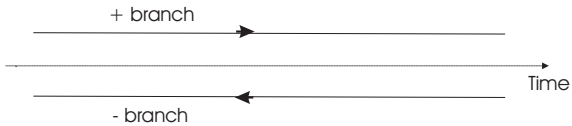}}
\caption{The Keldysh contour: Operators are ordered along the
contour
 with operators evaluated at later times acting on the left of operators evaluated at earlier times.
 Times on the (+) branch are always earlier than times on the (--) branch.}
\end{figure}
The non-equilibrium effect caused by the bias voltage $V$ can be
included in the density matrix. We assume that before the
backscattering Hamiltonian $H_{\rm bs}$ is turned on (at
$t_{0}\rightarrow-\infty$) the system has a well defined
non-equilibrium state determined by separate chemical potentials
for left and right-movers kept fixed by the chemical potentials of
the right and left electron reservoirs, respectively. The initial
density matrix therefore takes on the form \cite{PBW}
\begin{equation}
{\hat \rho}_{V}=\frac{1}{Z_{V}}e^{-\beta H_{V}},
\end{equation}
with $H_{V}=H_{\text{\textsc{swnt}}}-\mu_{R}N_{R}-\mu_{L}N_{L}$
and $\beta=1/k_{\rm B}T$ with $T$ the temperature, $k_{\rm B}$ the
Boltzmann constant and $Z_{V}={\rm Tr}[\exp(-\beta H_{V})]$ the
partition function. The equilibrium chemical potential is defined
as zero (a non-zero chemical potential can be taken into account
by the gate voltage) and $N_{R/L}=\int\,dx \,n_{R/L}(x)$. The bias
voltage is then related to the chemical potentials of left and
right-movers via $\mu_{R/L}=\pm eV/2$. As outlined in Ref.~11, it
is convenient to apply a unitary transformation $U_{V}$ such that
$U_{V}^{\dagger}H_{V}U_{V}=H_{\text{\textsc{swnt}}}+{\rm const}$..
This transforms the bias voltage from the density matrix (${\hat
\rho}_{V}\rightarrow{\hat \rho}_{0}$) into the backscattering
Hamiltonian $H_{\rm bs}$ which receives a time dependent phase
factor in the interaction picture governed by the shift
$\theta_{1}\rightarrow\theta_{1}-eVt$. In addition, the unitary
transformation transforms the observable according to $\hat{{\cal
O}}\rightarrow U_{V}^{\dagger}{\hat {\cal O}}U_{V}$. In the case
of the current operator this leads to the shift \cite{PBW}
$\hat{I}\rightarrow I_{0}+\hat{I}$. The average current can then
be written as
$\langle\hat{I}(x,t)\rangle=I_{0}+\langle\hat{I}_{V}(x,t)\rangle_{0}$.
Here, $I_{0}=4e^{2}V/h$ is the ideal current without
backscattering and $\hat{I}_{V}(x,t)$ gives rise to the
backscattered current
\begin{eqnarray}
\label{currentk1}
&&I_{B}(x,t)\equiv\langle\hat{I}_{V}(x,t)\rangle_{0}\nonumber\\
&& =\langle {\hat
T}_{K}\hat{I}_{K}(x,t)\,e^{-i\sum\limits_{r}r\int dt' H^{r}_{\rm
bs}(t')\bigr|_{\theta_{1}^{r}\rightarrow\theta_{1}^{r}-eVt'}}\rangle_{0}.
\end{eqnarray}
Here, we have introduced the Keldysh current operator
$\hat{I}_{K}(x,t)=(1/2)\sum_{r}\hat{I}^{r}(x,t)$, the time
ordering operator ${\hat T}_{K}$ along the Keldysh contour
depicted in Fig.~2 and $r=\pm$ which refers to fields defined on
the $\pm$-branch of that contour. In Eq.~(\ref{currentk1}) the
time dependence of all operators is due to
$H_{\text{\textsc{swnt}}}$ only and $\langle...\rangle_{0}={\rm
Tr}[{\hat \rho}_{0}...]$. A similar procedure can be performed for
the noise spectral density $S(x,\omega)=\int dt
e^{i\omega(t-t')}\,S(x;t,t')$, where the symmetrized
current-current correlator is $S(x;t,t')=(1/2)\langle\{\delta{\hat
I}(x,t),\delta{\hat I}(x,t')\}\rangle$ where $\{...\}$ denotes the
anticommutator, $\delta{\hat I}(x,t)={\hat I}(x,t)-\langle {\hat
I}\rangle$ and $\langle...\rangle={\rm Tr}[{\hat \rho}_{V}...]$
with the initial density matrix ${\hat \rho}_{V}$ discussed
before. Using again formally $\hat{I}(t)=I_{0}+\hat{I}_{V}(t)$ we
obtain $S(x,\omega)=(1/2)\int dt e^{i\omega(t-t')}\langle\{{\hat
I}_{V}(x,t),{\hat
I}_{V}(x,t')\}\rangle_{0}-2\pi\delta(\omega)I_{B}^{2}$. To lowest
order in the backscattering, the $\delta$-function contribution at
zero frequency can be neglected and we can therefore write the
current-current correlator as

\begin{eqnarray}
\label{keldyshnoise1}
 &&S(x;t,t')=\nonumber\\&&\langle {\hat T}_{K}{\hat
I}_{K}(x,t){\hat I}_{K}(x,t')\,e^{-i\sum\limits_{r}r\int dt''
H_{\rm bs}^{r}(t'')\bigr|_{\theta_{1}^{r}\rightarrow\theta_{1}^{r}-eVt''}}\rangle_{0}.\nonumber\\
\end{eqnarray}
The time ordered correlation functions can be conveniently
calculated by means of a functional integral approach discussed
next.

\subsection{The generating functional}

The statistical averages in Eqs.~(\ref{currentk1}) and
(\ref{keldyshnoise1}) are conveniently evaluated in terms of the
following generating functional $Z^{\eta}$
\begin{align}
\label{generfunctional}
 &Z^{\eta}=\prod_{a}\int\,{\cal
D}[\theta_{a}^{\pm}(t')\phi_{a}^{\pm}(t')]\nonumber\\
&\times\exp\left[iS_{0}-i\int dt'[H_{{\rm bs}}^{+}(t')-H_{{\rm
bs}}^{-}(t')]\bigr|_{\theta_{1}^{\pm}\rightarrow
\theta_{1}^{\pm}-eVt'}\right.\nonumber\\
 &\left.-i\int dt'\int
dx'\eta(x',t')\,\dot{\theta}_{1}(x',t')\right].
\end{align}
 We have performed the rotation to new fields
$\theta_{a}^{\pm}=\theta_{a}\pm i{\tilde{\theta}_{a}}/2$ which
allows the simple representation ${\hat
I}_{K}(x,t)=-e(2/\pi)\dot{\theta_{1}}(x,t)$. In
Eq.~(\ref{generfunctional}) we have introduced a source field
$\eta(x,t)$ which does not have a direct physical meaning but is
rather a convenient way to produce correlation functions via of
functional derivatives. The action $S_{0}$ describes the dynamics
induced by $H_{SWNT}$ only and is a quadratic form of the phase
fields $\theta_{a}(x,t),{\tilde \theta}_{a}(x,t)$ and
$\phi_{a}(x,t),{\tilde \phi}_{a}(x,t)$. The explicit form of
$S_{0}$ is presented in Appendix A. Here, we only give the
relevant correlation functions
\begin{align} \label{correlation}
C^{\theta\theta}_{a}(x,x';t)\equiv&\langle {\hat T}_{K}
\theta_{a}(x,t)\theta_{a}(x',0)\rangle_{0}\nonumber\\
=&\frac{1}{2}\langle\{{\theta}_{a}(x,t),{\theta}_{a}(x',0)\}\rangle_{0},
\end{align}
and the retarded functions
\begin{align}
\label{response} R^{\theta\theta}_{a}(x,x';t)\equiv&\langle {\hat
T}_{K}\theta_{a}(x,t){\tilde
\theta}_{a}(x',0)\rangle_{0}\nonumber\\
=&-i\Theta(t)\,\langle[{\theta}_{a}(x,t),{\theta}_{a}(x',0)]\rangle_{0},
\end{align}
and similar for $\phi_{a}$-correlations. Other combinations like
$\langle{\tilde \theta}_{a}(x,t){\tilde
\theta}_{a}(x',0)\rangle_{0}=\langle{\tilde \phi}_{a}(x,t){\tilde
\phi}_{a}(x',0)\rangle_{0}=0$.

\subsection{Shifted action}

It is advantageous to transform away the linear $\eta$-term in the
generating functional $Z^{\eta}$ by shifting the
$\theta_{1}$-fields such that in the new variables the linear term
in $\theta_{1}$ gets cancelled, whereas $S_{0}$ remains unchanged.
Since we have to perform such a transformation on the whole
action, including the backscattering contribution, the
$\eta$-source field will appear in the backscattering Hamiltonian
instead. This transformation we find to be
\begin{eqnarray}
\label{etatrafo} &&\theta_{1}(x,t)\rightarrow
\theta_{1}(x,t)\nonumber\\
 &&+\frac{1}{2\pi}\int dx'\int
d\omega\,\omega\,e^{-i\omega
t}\,\eta(x',\omega)\,C_{1}^{\theta\theta}(x,x';\omega),\nonumber\\
&&\nonumber\\
&&{\tilde\theta}_{1}(x,t)\rightarrow
{\tilde\theta}_{1}(x,t)\nonumber\\
 &&+\frac{1}{2\pi}\int dx'\int
d\omega\,\omega\,e^{-i\omega
t}\,\eta(x',\omega)\,R_{1}^{\theta\theta}(x',x;-\omega).\nonumber\\
\end{eqnarray}
 Since the action $S_{0}$ couples $\phi_{1}$ and
$\theta_{1}$ (see Appendix A), $\phi_{1}$ gets also transformed.
However, its transformation is not needed here since
$\phi_{1}$-terms are absent in $H_{\rm bs}$ [see
Eq.~(\ref{backh})] which states that the total charge is conserved
in the backscattering process. In the new variables the generating
functional becomes
\begin{eqnarray}
\label{Zform}
 &&Z^{\eta}=e^{-\frac{1}{4\pi}\int
d\omega\omega^{2}\int dx'\int dx''
\eta(x',\omega)^{*}C_{1}^{\theta\theta}(x',x'';\omega)\eta(x'',\omega)}\nonumber\\
&&\times\langle
e^{-i\,\int\,dt'(\overrightarrow{H}_{bs}^{+}-\overrightarrow{H}_{bs}^{-})}\rangle_{0},
\end{eqnarray}
where $\eta(x,\omega)^{*}=\eta(x,-\omega)$ and we used the
abbreviation $\langle...\rangle_{0}=\prod_{a}\int {\cal
D}[\theta_{a}^{\pm}\phi_{a}^{\pm}]...\exp[iS_{0}]$. The arrow
$\overrightarrow{}$ in Eq.~(\ref{Zform}) depicts the shift of
$\theta_{1}^{\pm}$ via Eq.~(\ref{etatrafo}) and the effect of the
applied voltages, explicitly
$\theta_{1m}^{\pm}(t)\rightarrow\theta_{1m}^{\pm}(t)+(1/2\pi)\int
d\omega\int dx'\omega e^{-i\omega
t}\eta(x',\omega)[C_{1}^{\theta\theta}(x_{m},x';\omega)\pm\frac{i}{2}R_{1}^{\theta\theta}(x',x_{m};-\omega)]-eVt-V_{\rm
g}x_{m}$. The generating functional Eq.~(\ref{Zform}) is the
starting point for calculating any order of current-current
correlation functions for a general measurement position $x$.
\section{dc current}
In this section we derive and analyze the  dc current
$\langle\hat{I}\rangle\equiv I=I_{0}+I_{B}$ and the conductance
$G=dI/dV$ at finite temperatures. We will first present the
general result for arbitrary bias voltage, gate voltage and
temperature to leading order in the backscattering of the two
barriers followed by analytical approximations and a discussion of
the results. The backscattered current is given in terms of the
generating functional Eq.~(\ref{Zform}) by
\begin{equation}
I_{B}(x,t)=-i\frac{2e}{\pi}\frac{\delta}{\delta\eta(x,t)}\,Z^{\eta}\biggr|_{\eta=0}.
\end{equation}
The actual derivation of the result is straightforward but
lengthy. Some of the methods  and intermediate results are
presented in Appendix B. The final result for the current can be
written as $I=I_{0}+I_{B}^{\rm in}+I_{B}^{\rm co}$, explicitly:
\begin{eqnarray}
\label{backscatteredcond}
 &&I=I_{0}[1+U^{{\rm in}}\frac{1}{v}\int
d\tau\,
e^{\textbf{C}_{11}(\tau)}\sin[\textbf{R}_{11}(\tau)/2]\sin(v\tau)\nonumber\\
&&+U^{{\rm co}}\frac{1}{v}\int d\tau\,
e^{\textbf{C}_{12}(\tau)}\sin[\textbf{R}_{12}(\tau)/2]\sin(v\tau)],
\end{eqnarray}
with effective backscattering strengths $U^{\rm
in}=\sum_{m=1,2}U_{m}^{\rm in}$, where
\begin{displaymath}
U_{m}^{\rm in}=(4\pi
t_{F}^{2}/\hbar^2)e^{-\textbf{C}_{11}(0)}\,\sum\limits_{ij}(u_{m}^{ij})^2,
\end{displaymath}
and
\begin{displaymath}
U^{\rm co}=(8\pi
t_{F}^{2}/\hbar^2)e^{-\textbf{C}_{11}(0)}\,\sum\limits_{ij}u_{1}^{ij}u_{2}^{ij}\cos(V_{\rm
g}L+2\Delta_{ij}).
\end{displaymath}
We have introduced the dimensionless time
$\tau=t/t_{\text{\textsc{f}}}$ with
$t_{\text{\textsc{f}}}=L/v_{\text{\textsc{f}}}$ the
non-interacting traversal time of the SWNT as well as the
dimensionless voltage $v=eVt_{\text{\textsc{f}}}/\hbar$. Note that
the dc current is independent of the measurement point $x$ and
time $t$. Each backscattering event involves a combination of the
total charge mode ($\theta_{1}\theta_{1}$-correlations) and the
three non-interacting modes  ($\theta_{a}\theta_{a}$- or
$\phi_{a}\phi_{a}$-correlations, $a=2,3,4$). Therefore,
$\textbf{C}_{mm'}(t)=C^{{\rm I}}_{mm'}(t)+3\,C^{{\rm F}}_{mm'}(t)$
and a similar definition holds for $\textbf{R}_{mm'}(t)$
$(C\rightarrow R)$. The superscripts ${\rm I}$ and ${\rm F}$ refer
to interacting ($g<1$) and free (non-interacting, i.e. $g=1$),
respectively. Eq.~(\ref{backscatteredcond}) is consistent with the
conductance formula derived in Ref.~11 up to the scattering phases
$\Delta_{ij}$ which have been neglected previously. Physically,
the term in Eq.~(\ref{backscatteredcond}) proportional to $U^{\rm
in}$ describes the incoherent addition of two barriers whereas the
term proportional to $U^{\rm co}$ describes the quantum mechanical
interference between backscattering events of different barriers
(1 or 2). Note that the interference term can be modulated by the
gate voltage $V_{\rm g}$. In addition, different scattering phases
$\Delta_{ij}$ for intraband ($i=j$) and interband ($i\neq j$)
processes lead to a reduction of the FP-amplitude (see also
Ref.~8). A general analytical form of the conductance seems
difficult to derive and we have to rely on numerical integration
of Eq.~(\ref{backscatteredcond}). The main physics can
nevertheless be understood in terms of the correlation- and
retarded functions to be discussed next.

\subsection{Retarded and correlation functions}
Here, we present the results for the retarded functions and
correlation functions which are carefully  derived in Appendix C.
In general, the retarded functions can be written  as
$R_{mm'}^{{\rm I(F)}}(\tau)=\theta(\tau)[r_{mm'}^{\rm
I(F)}(\tau)-r_{m'm}^{\rm I(F)}(-\tau)]$, where $r_{mm'}^{\rm
I(F)}(\tau)$ are the Fourier transforms of the retarded Green's
functions in frequency space using a high-energy cut-off function
$\exp(-|\omega|/\omega_{0})$. For the interacting (I) correlations
at the same barriers we obtain
$r_{11}^{\rm{I}}(\tau)=r_{22}^{\rm{I}}(\tau)$ with

\begin{align}
\label{reta11} r_{11}^{{\rm I}}(\tau)
=&-\frac{\pi}{2}(1-\gamma)\nonumber\\
 \times&\left\{\Theta_{\alpha}(\tau)+\frac{1+\gamma}{\gamma}\sum\limits_{k=1}^{\infty}\gamma^{2k}\Theta_{\alpha}(\tau-2kg)\right\}.\nonumber\\
\end{align}
Here, the interaction parameter $g$ is introduced via
$\gamma=(1-g)/(1+g)$ which can be interpreted as the reflection
coefficient for an incoming charge flux traversing the
reservoir-nanotube interface \cite{SafiSchulz}. For the non-local
correlations we obtain  $r_{12}^{{\rm I}}(\tau)=r_{21}^{{\rm
I}}(\tau)$ with

\begin{equation}
\label{reta12} r_{12}^{{\rm
I}}(\tau)=-\frac{\pi}{2}(1-\gamma^2)\sum\limits_{k=0}^{\infty}\,\gamma^{2k}
\Theta_{\alpha}[\tau-(2k+1)g].
\end{equation}
The smeared step function is defined as $
\Theta_{\alpha}(\tau)=(1/\pi)\arctan\left(\frac{\tau}{\alpha}\right)+1/2
$ where $\alpha=(t_{\text{\textsc{f}}}\omega_{0})^{-1}$. The high
energy cut-off of the theory is $\epsilon_{0}=\hbar\omega_{0}\sim
1 {\rm eV}$ which is the bandwidth of the SWNT. In all plots we
will fix $\alpha=0.001$ and $v_{\text{\textsc{f}}}=8\times 10^{5}$
m/s which corresponds to a nanotube length of $L\sim$ 527 nm
relevant for existing experiments on two-terminal ballistic
transport \cite{HongkunPark, Kong,Kim}. The non-interacting
retarded functions $R_{mm'}^{{\rm F}}(\tau)$ are obtained from the
interacting ones by setting $g=1$. The correlation function we
decompose into a zero temperature part plus the finite temperature
correction as
\begin{equation}
C_{mm'}^{{\rm
I(F)}}(\tau)=C_{mm'}^{\rm{I(F)0}}(\tau)+C_{mm'}^{\rm{I(F)T}}(\tau).
\end{equation}
The interacting correlation functions at zero temperature are
given as $C_{11}^{\rm{I0}}(\tau)=C_{22}^{\rm{I0}}(\tau)$ with
\begin{multline}
\label{T1}
C_{11}^{\rm{I0}}(\tau)=-\frac{1-\gamma}{4}\biggl\{\ln(\alpha^2+\tau^2)\\
+\frac{1+\gamma}{2\gamma}\sum\limits_{k=1}^{\infty}\gamma^{2k}\sum\limits_{r=\pm}\,\ln\left[\alpha^2+(\tau+r2kg)^2\right]\biggr\}.
\end{multline}
For the cross-terms we obtain
$C_{12}^{\rm{I0}}(\tau)=C_{21}^{\rm{I0}}(\tau)$ with
\begin{multline}
\label{T2}
C_{12}^{\rm{I0}}(\tau)=-\frac{1-\gamma^2}{8}\\
\times\sum\limits_{k=0}^{\infty}\gamma^{2k}\sum\limits_{r=\pm}\,\ln\left[\alpha^2+(\tau+r(2k+1)g)^2\right].\\
\end{multline}
In Eqs. (\ref{T1}) and (\ref{T2}) we have dropped a
$\tau$-independent and $mm'$-independent constant which does not
contribute to the relevant combination ${\bf C}_{mm'}(\tau)-{\bf
C}_{11}(0)$. In the finite temperature part
$C_{mm'}^{\rm{IT}}(\tau)$ the high energy cut-off $\epsilon_{0}$
can be sent to infinity ($\alpha\rightarrow 0$) as the cut-off is
now played by the finite temperature (the result for finite
$\alpha$ is presented in Appendix C). We obtain
\begin{multline}
\label{c11}
C_{11}^{\rm{IT}}(\tau)=\frac{1-\gamma}{2}\ln\left[\frac{\pi\Xi
\tau}{\sinh(\pi\Xi \tau)}\right]\\
+\frac{1-\gamma^2}{4\gamma}\sum\limits_{k=1}^{\infty}\gamma^{2k}\sum\limits_{r=\pm}\,\ln\left[\frac{\pi\Xi(\tau+r2kg)}{\sinh[\pi\Xi(\tau+r2kg)]}\right],\\
\end{multline}
and the same for $C_{22}^{\rm{IT}}(\tau)$. For the interference
term we find
\begin{multline}
\label{c12}
C_{12}^{\rm{IT}}(\tau)=\frac{1-\gamma^2}{4}\\
\times\sum\limits_{k=0}^{\infty}\,\gamma^{2k}\sum\limits_{r=\pm}\,\ln\left[\frac{\pi\Xi(\tau+r(2k+1)g)}{\sinh[\pi\Xi(\tau+r(2k+1)g)]}\right],\\
\end{multline}
and the same for $C_{21}^{\rm{IT}}(\tau)$. In both correlation
functions the dimensionless temperature is $\Xi=k_{\rm
B}Tt_{\text{\textsc{f}}}/\hbar$. The non-interacting functions
$C_{mm'}^{{\rm F}}$ are obtained by setting $g=1$ in $C_{mm'}^{\rm
I}$. We note that all correlation- and retarded functions agree
with the zero temperature results given in Ref.~11 in the limit
$\alpha\rightarrow 0$. However, we note that a finite cut-off is
crucial when doing the time integral in
Eq.~({\ref{backscatteredcond}}) for the case of $g=1$.
\begin{figure}[h]
\begin{center}
\hbox{\resizebox{6.5cm}{!}{\includegraphics{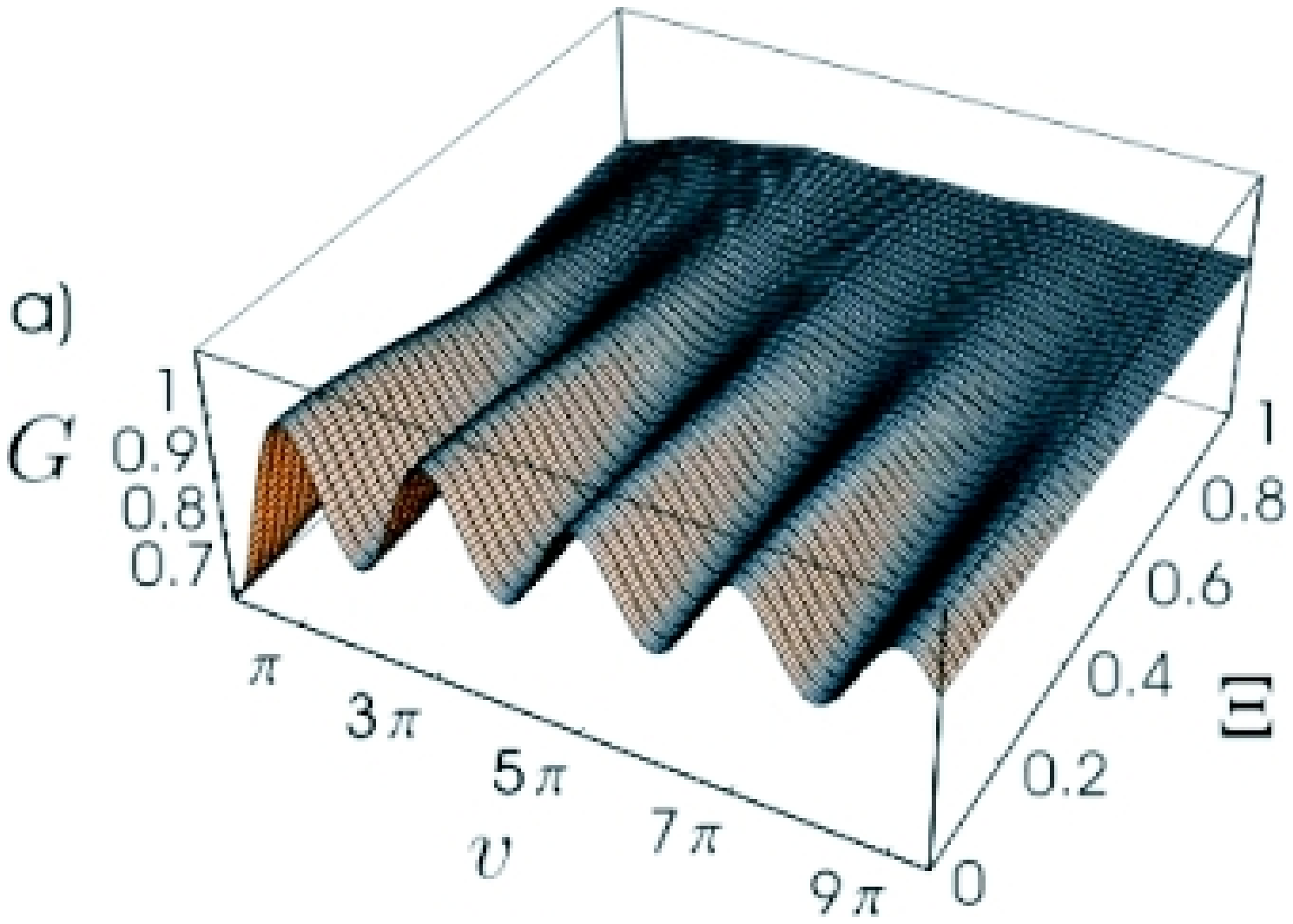}}}
\end{center}
\begin{center}
\hbox{\resizebox{4.2cm}{!}{\includegraphics{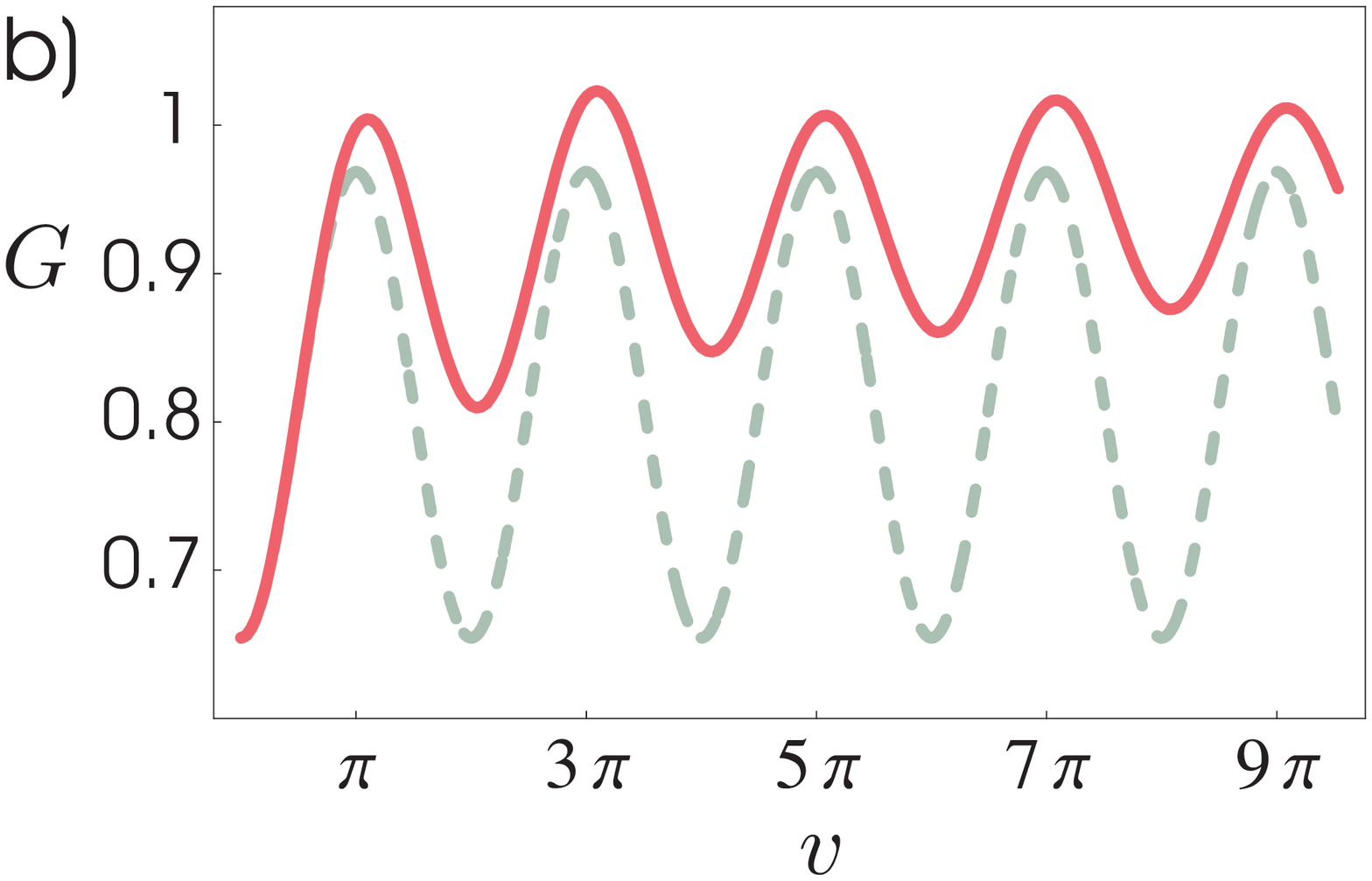}}\,\,\resizebox{4.2cm}{!}{\includegraphics{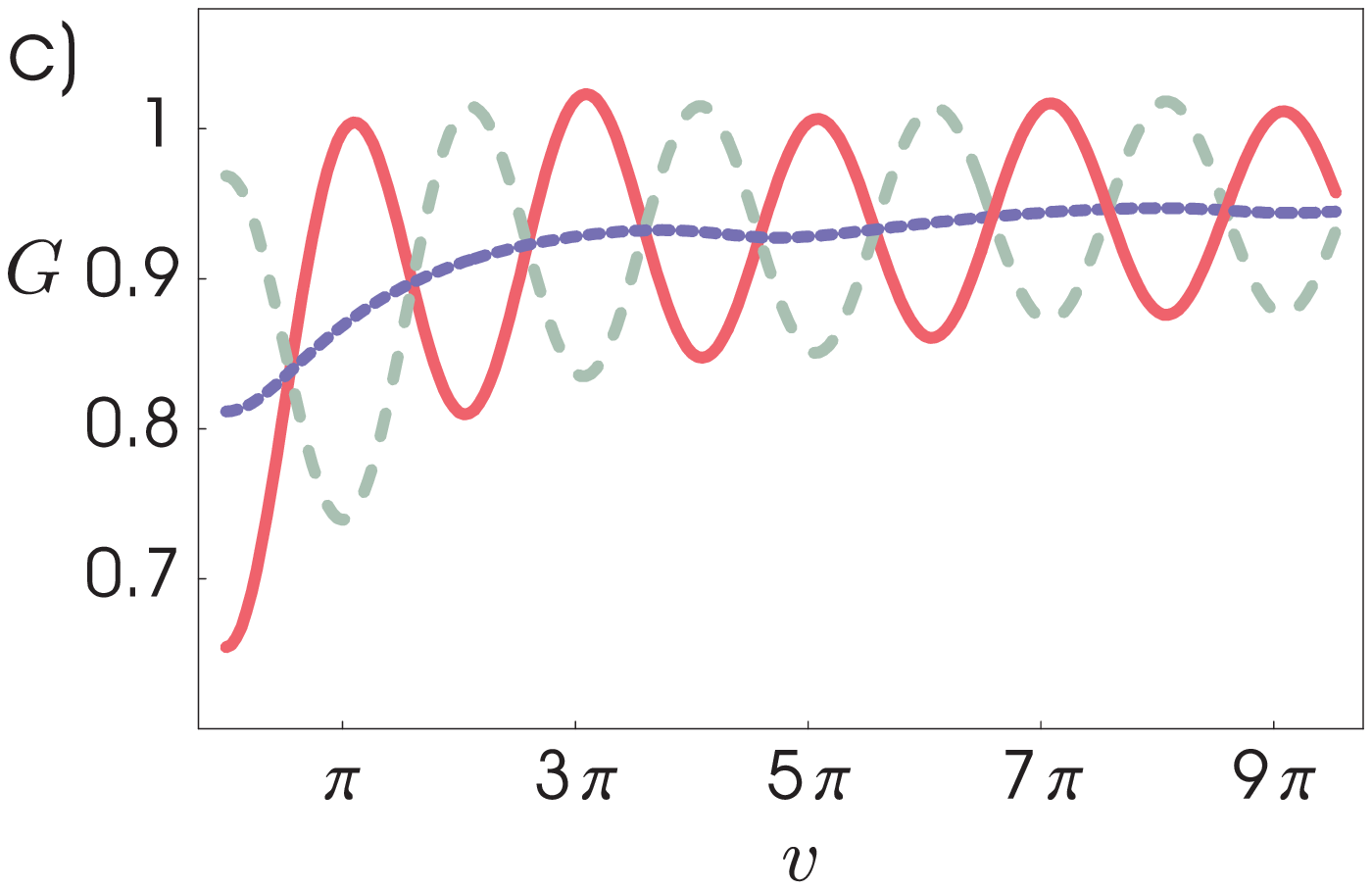}}}
\end{center}
\caption{(color online). Conductance plots: in Fig.~a) we show the
conductance as a function of bias voltage $v$ and temperature
$\Xi$ both in units of the non-interacting level spacing
$\hbar/t_{\text{\textsc{f}}}$ for the strongly correlated case
$g=0.23$. The backscattering coefficients are taken at zero
temperature as $U^{\rm in}=0.12$, $U^{\rm co}=0.1$. In Fig.~b) we
compare the conductance at zero temperature for different
interaction strengths: $g=0.23$(red), $g=1$(dashed). Fig.~c) is
devoted to the study of the gate voltage dependence for $g=0.23$:
we have chosen $U^{\rm in}=0.12$, $U^{\rm co}=0.1$(red), $U^{\rm
in}=0.12$, $U^{\rm co}=-0.1$(dashed), $U^{\rm in}=0.12$, $U^{\rm
co}=0$(points). }
\end{figure}
\subsection{Analytical results}
In this subsection we provide an analytical approximation of
$I_{B}^{\rm in}$  in the regime where the bias voltage and/or
temperature are large compared to the interacting level spacing
$\hbar/t_{c}$ where $t_{c}=t_{\text{\textsc{f}}}g$ is the charge
traversal time along the SWNT. In the non-interacting case $g=1$,
we can calculate $I_{B}$
 analytically without approximations, including the interference term $I_{B}^{\rm
 co}$.

Since the correlation time for the backscattering processes is
given by $\hbar/eV$ or $\hbar/k_{\rm B}T$ the multiple reflection
terms in the retarded and correlation functions [Eqs.~(15)-(21)]
are not resolved as the traversal time $t_{c}$ is too large. We
then take only the $k=0$ contribution in the retarded function
[Eq.~(\ref{reta11})] and set $\tau$ to zero in the $k\geq 1$ terms
in the correlation functions [Eqs.~(\ref{T1}) and (\ref{c11})].
Note that the time $t$ has to be considered as still larger than
the cut-off time $\hbar/\epsilon_{0}$.
\begin{figure}[h]
\centerline{\includegraphics[width=6.0cm]{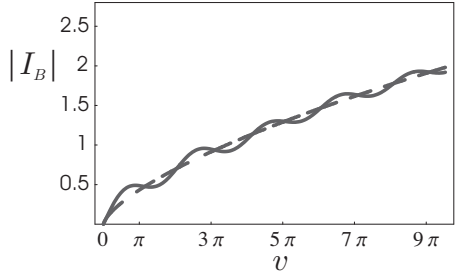}}
\caption{Comparison of backscattered current $I_{B}$ in units of
$G_{0}\hbar/t_{\text{\textsc{f}}}e$ at zero temperature (black)
with the approximate formula Eq.~(\ref{explicitform}) (dashed)
showing the power-law $I_{B}\propto V^{1-\gamma/2}$. In both cases
we use $U^{\rm in}=U^{\rm co}=0.1$ and $g=0.23$.}
\end{figure}
Explicitly, we consider the regime ${\rm max}(eV,k_{\rm B}T)\gg
\hbar/t_{c}$ where the incoherent portion $I_{B}^{\rm in}$  of the
backscattered current becomes proportional to the integral
\begin{equation}
I_{B}^{\rm in}\propto\sum\limits_{r=\pm}r\int
d\tau\frac{\sin(v\tau)}{\sinh[\pi\Xi(r \tau+i\alpha)]^{2\nu}}.
\end{equation}
This integral can be expressed in terms of standard functions with
the result (in the limit $\alpha\rightarrow 0$)
\begin{multline}
\label{fullT} I_{B}^{\rm in}=
-\frac{4e}{h}\sum\limits_{ij}\bigl[(u_{1}^{ij})^2+(u_{2}^{ij})^2\bigr]k_{\rm B}T\sinh\left(\frac{eV}{2k_{\rm B}T}\right)\\
\times\frac{(2\pi/\epsilon_{0})^2}{\Gamma(2-\gamma/2)}\left(\frac{2\pi
k_{\rm
B}T}{\epsilon_{0}}\right)^{-\gamma/2}\left|\Gamma\left(1-\frac{\gamma}{4}+i\frac{eV}{2\pi
k_{\rm B}T}\right)\right|^{2}.
\end{multline}
Note that for  $g=1$ the temperature dependence drops out and the
current is only depending on the bias voltage. This is only true
if the transmission is energy independent which is the case for
the incoherent contribution $I_{B}^{\rm in}$.
\begin{figure}[h]
\centerline{\includegraphics[width=5.9cm]{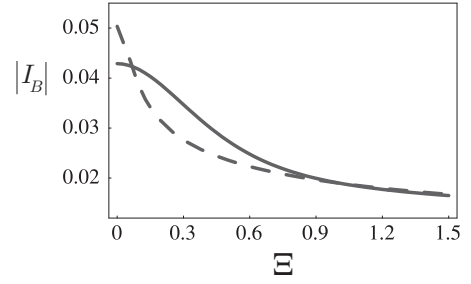}}
\caption{Backscattered current $I_{B}$ in units of
$G_{0}\hbar/t_{\text{\textsc{f}}}e$ at $v\sim 0.137$ as a function
of dimensionless temperature $\Xi$. The continuous line is
obtained from numerical integration of
Eq.~(\ref{backscatteredcond}), whereas the dashed line is the
approximated incoherent contribution of $I_{B}$ given in
Eq.~(\ref{fullT}). As expected, they agree for temperatures larger
than the non-interacting level spacing, i.e $\Xi>1$. For both
curves we use $U^{\rm in}=U^{\rm co}=0.1$ and $g=0.23$.}
\end{figure}
In the high bias regime $eV\gg k_{\rm B}T$ we obtain the power-law
scaling on bias voltage
\begin{multline}
\label{explicitform}
I_{B}^{{\rm in}}=-\frac{2e^2}{h}V\sum\limits_{ij}\bigl[(u_{1}^{ij})^2+(u_{2}^{ij})^2\bigr]\\
\times\frac{(2\pi/\epsilon_{0})^2}{\Gamma(2-\gamma/2)}\left(\frac{eV}{\epsilon_{0}}\right)^{-\gamma/2}e^{-eV/\epsilon_{0}}.
\end{multline}
The interference contribution proportional to $U^{\rm co}$
describes the FP-interference with two frequencies coming from the
total charge mode with velocity $v_{c}$ and three non-interacting
modes with velocities $v_{\text{\textsc{f}}}$. In general, such
integrals are not straightforward to calculate analytically even
in the high energy regime unless we set $g=1$.

When $g=1$, we can calculate the backscattered current
analytically for all temperatures and bias voltages. The
FP-interference contribution $I_{B}^{\rm co}$ then becomes
proportional to the integral
\begin{equation}
I_{B}^{{\rm co},g=1} \propto\pi\Xi\int
d\tau\frac{e^{i\Omega\tau}-e^{-i\Omega\tau}}{\prod\limits_{r=\pm}\sinh(\tau+r\pi\Xi+i\alpha)},
\end{equation}
where $\Omega=v/\pi\Xi$ and $\alpha\rightarrow 0^{+}$. This
integral has simple poles (for $\Xi\neq 0$) for
$\tau=-r\pi\Xi+in\pi-i\alpha$ where $n=0,\pm1,\pm2...$. If
$\Omega>0$ we can close the contour in the upper half of the
complex plane associated with $e^{i\Omega\tau}$ thereby picking up
poles for $n\geq 1$, and in the lower half-plane associated with
the  term  $e^{-i\Omega\tau}$ and picking up poles for $n\leq 0$.
All poles except the one for $n=0$ cancel when combining the two
contributions. Adding the $g=1$ contribution from
Eq.~(\ref{fullT}) (or Eq.~(\ref{explicitform}) in the limit
$\epsilon_{0}\rightarrow \infty$) the result is
\begin{multline}
\label{Tg1}
I_{B}^{g=1}=-\frac{2e^2}{h}\left(\frac{2\pi}{\epsilon_{0}}\right)^{2}\biggl\{\sum\limits_{ij}\bigl[(u_{1}^{ij})^{2}+(u_{2}^{ij})^{2}\bigr]V\\
+\sum\limits_{ij}2u_{1}^{ij}u_{2}^{ij}\cos(V_{\rm g}L+2\Delta_{ij})\\
\times\frac{(2\pi k_{\rm B}T/e)}{\sinh(2\pi k_{\rm
B}Tt_{\text{\textsc{f}}}/\hbar)}\sin\left(\frac{eVt_{\text{\textsc{f}}}}{\hbar}\right)\biggr\}.
\end{multline}
We note that temperature suppresses the FP-interference
exponentially if $k_{\rm B}Tt_{\text{\textsc{f}}}/\hbar\gg 1$,
i.e. if the temperature is much larger than the level spacing.
\subsection{Physical interpretation of dc current results}
In Section B we derived the backscattered current and conductance
as a function of bias voltage, gate voltage and temperature.

First, we discuss the incoherent contribution of $I_{B}$ which is
dominant at large bias voltages. As the energy scale at which the
system is probed exceeds the interacting charge mode level spacing
$\hbar v_{c}/L$, the TLL-correlations become apparent and our
asymptotic formula Eq.~(\ref{fullT}) applies approximately (see
Figs.~4 and 5). At large bias voltages (and small temperatures) we
observe the characteristic power-law in Eq.~(\ref{explicitform}).
At this energy scale, the $U^{\rm in}$-term (incoherent part) does
not resolve the multiple reflections of the charge mode
originating from the inhomogeneity of $g$ at the boundaries
between SWNT and reservoirs because the traversal time $t_{c}$
becomes larger than the coherence time of electron wave packets
given by $\hbar/eV$ or $\hbar/ k_{\rm B}T$. In this case, only the
$k=0$ term in
 retarded-  and correlation functions contributes significantly. The
strength of charge mode ($a=1$) correlations relative to the
non-interacting modes $a=2,3,4$ is then given as $1-\gamma$ where
$\gamma=(1-g)/(1+g)$. This is apparent from the formula for the
retarded function Eq.~(\ref{reta11}) or in the correlation
functions Eqs.~(\ref{T1}) and (\ref{c11}). This relative factor
$1-\gamma$ is the effective TLL-parameter $g_{{\rm eff}}$ at the
boundary connecting a Fermi liquid system (metal reservoir) with
an {\it infinite} TLL-system \cite{Sandler}. We therefore conclude
that at high voltages (or at high temperatures), a charge
$g_{\rm{eff}}e$ gets locally backscattered. Similar, the
interference term proportional to $U^{\rm co}$ involves the
combination $(1-\gamma)(1+\gamma)=1-\gamma^2$ which is the product
of two backscattering events at spatially separated places [see
Eqs.~(\ref{reta12}), (\ref{T2}) and (\ref{c12})]. The
$(1+\gamma)$-factor appears because the two backscattered charges
can only interfere after the backscattered charge at the second
barrier traverses the SWNT and has to be transmitted to the left
contact with an additional factor $1+\gamma$ on the way. In
general, the power-law behavior of transport can be understood as
an energy-dependent renormalization of the bare backscattering
amplitudes $u_{m}^{ij}$ due to electron-electron interactions. It
is a well known fact that a weak backscatterer grows strong as one
approaches low energies, eventually going into the tunneling
regime. This is signalled by a divergent power-law at small
energies \cite{KaneFisher92}. In our calculation we take into
account the finite-size effect of the interacting region and
therefore will not encounter this divergence as the power-law is
only valid above the charge mode level spacing. For sufficiently
small bare backscattering amplitudes, the perturbative approach
presented in this work is therefore valid at all energy scales.
Indeed, at energies below the interacting level spacing, the
coherence time of electron wave packets becomes much larger than
the traversal time and eventually all multiple reflections
contribute. Formally this limit corresponds to $L\rightarrow 0$ or
$\tau\rightarrow \infty$ in the time integrals of
Eq.~(\ref{backscatteredcond}) where we can sum up all $k$-terms
and getting back the non-interacting functions. Therefore, the
backscattered current is linear as $V\rightarrow 0$.

 The interference contribution proportional to
$U^{\rm co}$ shows not only a reduction of the amplitude when
sweeping the bias voltage to higher values but is in principle
capable of showing FP-oscillations containing two frequencies,
namely $v_{c}/L$ coming from the charge mode $a=1$, and
$v_{\text{\textsc{f}}}/L$ defining the frequency of the
non-interacting modes $a=2,3,4$. However, the visibility of the
interacting mode is in general much less pronounced than the
non-interacting modes as can be seen in Figs.~3b) and 3c). The
reason is two-fold: First, all backscattering processes involve
three non-interacting modes and only one interacting mode.
Therefore, the contribution of the total charge mode $a=1$ is less
pronounced. Second, the interacting mode contribution is further
reduced  by the smallness of $g_{{\rm eff}}$ which enters as a
prefactor in the retarded  as well as correlation functions at
high energies (i.e. small times $t$). At small energies (i.e.
large times $t$), all multiple reflections contribute, and
therefore the weight of charge mode oscillations increases, but
then the charge mode behaves effectively as a non-interacting mode
where the separation of velocities is absent.

A finite temperature, besides triggering the TLL-effect [see
Eq.~(\ref{fullT})], has an additional impact on suppressing the
FP-oscillation amplitude which is also the case for the
non-interacting system [see Eq.~(\ref{Tg1})]. This is caused by
the temperature induced smearing of the reservoirs Fermi
functions. But note that this suppression becomes exponential at
large temperatures whereas the TLL-effect is power-law.

In contrast to the bias voltage or temperature, the gate voltage
does not enter as a power-law and leads essentially to a periodic
modulation of  $U^{\rm co}$, see Eq.~(\ref{backscatteredcond}). In
Ref.~11 it was proposed that changing the gate voltage $V_{\rm g}$
allows to tune the strength of ordinary FP-oscillations ($U^{\rm
co}$-term) relative to the incoherent contribution ($U^{\rm
in}$-term) which is less sensitive (through a weak gate voltage
dependence of $g$). The oscillations contained in the incoherent
contribution is an interference effect due to a single impurity:
the backscattered charge at the impurity can interfere with the
{\it momentum-conserving} reflections due to the finite size of
the interacting region. Although such a dependence on the gate
voltage is expected, we note that the
 oscillations in the incoherent $U^{\rm in}$-term survive
only at small voltages $eV\lesssim \hbar
v_{\text{\textsc{f}}}/Lg$. This requires that the backscattering
must be very small in order for the perturbative treatment to be
valid. In contrast, the ordinary oscillations ($U^{\rm co}$-term)
due to two barriers are more stable towards higher voltages. We
will see that the distinction between the ordinary FP-oscillations
and the oscillations due to the finite-size effect of the
interacting region is much more apparent in the
frequency-dependent shot noise.
%
%
\section{Current noise}
The current noise $S(x,\omega)$ can be written in terms of the
generating functional Eq.~(\ref{Zform}) as
\begin{equation}
\label{noise1} S(x,\omega)=-\frac{8}{\pi}e^2\int
d\omega'\frac{\delta^2}{\delta\eta(x,\omega)^{*}\delta\eta(x,\omega')}Z^{\eta}\biggr
|_{\eta=0}.
\end{equation}
It is obvious from the general form of the shifted generating
functional  Eq.~(\ref{Zform}) that we can write the noise as
\begin{equation}
S(x,\omega)=S^{0}(x,\omega)+S_{I}(x,\omega),
\end{equation}
where $S^{0}(x,\omega)$ is the noise in the absence of
backscattering, and the impurity noise $S_{I}(x,\omega)$ is the
contribution due to electron backscattering at the SWNT-metal
reservoir interfaces. We first give the general result of current
noise. This is followed by a discussion of the low-frequency noise
and the Fano factor $F=S/eI$ relevant for existing experiments
\cite{Kim}. We then provide an analytical formula for the
high-frequency impurity noise for general interaction strength $g$
in terms of the incoherent contribution (only $U^{\rm
in}$-contributions) which is the dominant source of noise at high
energies. The general numerical evaluation including the
FP-interference is presented in Figs.~6-9. In the non-interacting
limit $g=1$, we can calculate the noise analytically.

The result for the current noise in the absence of backscattering
is
\begin{equation}
\label{cleannoisemain}
S^{0}(x,\omega)=G_{0}\omega\coth\left(\frac{\beta\omega}{2}\right){\rm
Re}\,\sigma_{0}(x,x;\omega),
\end{equation}
where ${\rm Re}$ means real-part, and we have introduced the
dimensionless conductivity \cite{conductivity} of the clean system
without the backscattering
$\sigma_{0}(x,y;\omega)=(2i\omega/\pi)R_{1}^{\theta\theta}(x,y;\omega)$.
In the following, we will discuss the frequency dependence of the
impurity noise $S_{I}$ and refer for simplicity of the discussion
to the situation where $\sum_{ij}u_{1}^{ij}u_{2}^{ij}\sin(V_{\rm
g}L+2\Delta_{ij})=0$ which can always be reached by tuning the
gate voltage. We present the general result in Appendix B. Then,
$U^{\rm co }\propto\sum_{ij}u_{1}^{ij}u_{2}^{ij}\cos(V_{\rm
g}L+2\Delta_{ij})$ will be maximal for this particular gate
voltage. We split the impurity noise in an incoherent part plus a
coherent part,
 $S_{I}(x,\omega)=S_{I}^{\rm in}(x,\omega)+S_{I}^{\rm
co}(x,\omega)$. In units of $G_{0}\hbar/t_{\text{\textsc{f}}}$ we
obtain
\begin{widetext}
\begin{align}
\label{sw2main} S_{I}^{\rm in}(x,\omega)=
&-\frac{et_{\text{\textsc{f}}}}{2\hbar G_{0}}\,\sum\limits_{r=\pm}\coth\left(\frac{v+r{\tilde \omega}}{2\Xi}\right)\,\sum\limits_{m}\,|\sigma_{0}(x,x_{m};\omega)|^2I_{Bm}^{{\rm in}}(v+r{\tilde \omega})\nonumber\\
&-2\sum\limits_{m}U_{m}^{{\rm in}}\,\coth\left(\frac{{\tilde
\omega}}{2\Xi}\right){\rm Re}\,\sigma_{0}(x,x_{m};\omega)\,{\rm
Im}\left[\sigma_{0}(x,x_{m};\omega)\,\int d\tau\, e^{{\bf
C}_{11}(\tau)}\sin[{\bf R}_{11}(\tau)/2]\left(1-e^{i{\tilde
\omega} \tau}\right)\cos(v\tau)\right],\nonumber\\
&\nonumber\\
&\nonumber\\
 S_{I}^{\rm
co}(x,\omega)=&-\frac{et_{\text{\textsc{f}}}}{2\hbar G_{0}}\,{\rm
Re}\left\{\sigma_{0}(x,x_{1};\omega)^{*}\sigma_{0}(x,x_{2};\omega)\right\}\sum\limits_{r=\pm}\coth\left(\frac{v+r{\tilde \omega}}{2\Xi}\right)\,I_{B}^{{\rm co}}(v+r{\tilde \omega})\nonumber\\
&-U^{\rm{co}}\,\coth\left(\frac{{\tilde
\omega}}{2\Xi}\right)\sum\limits_{mm'}\,{\rm
Re}\,\sigma_{0}(x,x_{m};\omega)\nonumber\\
&\times{\rm Im}\left\{\sigma_{0}(x,x_{m'};\omega)\int
d\tau\,e^{{\bf C}_{12}(\tau)}\sin[{\bf
R}_{12}(\tau)/2]\left(\delta_{mm'}-(1-\delta_{mm'})e^{i{\tilde
\omega} \tau}\right)\cos(v\tau)\right\}.
\end{align}
\end{widetext}
In Eq.~(\ref{sw2main}) we need the charge conductivity
$\sigma_{0}(x,x_{m};\omega)$ connecting the impurity positions
$x_{m}=\pm L/2$ with the point of measurement which we choose to
be in the right lead, i.e. $x\geq L/2$ (the result for $x$ in the
left lead is easily obtained from Eq.~(\ref{retardedfutext}) by
$x_{1}\leftrightarrow x_{2}$ and $x\rightarrow -x$). In this case
we get for the retarded function
\begin{multline}
\label{retardedfutext}
R_{1}^{\theta\theta}(x,x_{m};\omega)=-\frac{i\pi}{2\omega}(1-\gamma)\frac{e^{i\omega(\frac{x}{L}-\frac{1}{2})t_{\text{\textsc{f}}}}}{1-\gamma^2e^{i2\omega
t_{c}}}\\
\times\left(e^{i\omega(\frac{1}{2}-\frac{x_{m}}{L})t_{c}}+\gamma
e^{i\omega(\frac{3}{2}+\frac{x_{m}}{L})t_{c}}\right).
\end{multline}
Note that
$R_{1}^{\theta\theta}(x,x_{m};\omega)^{*}=R_{1}^{\theta\theta}(x,x_{m};-\omega)$.
In Eq.~(\ref{sw2main}) we have introduced the dimensionless
frequency ${\tilde \omega}=\omega t_{\text{\textsc{f}}}$ and the
decomposition $I_{B}^{\rm in}=\sum_{m=1,2}I_{Bm}^{\rm in}$.
\subsection{Low-frequency noise}
In this subsection we investigate the small frequency limit
$\omega\rightarrow 0$ of noise.
\subsubsection{Noise in the absence of backscattering}
We first consider Eq.~(\ref{cleannoisemain}) in the limit of small
$\omega$ where $S^{0}(x,\omega)\propto
 \coth(\beta\omega/2)\omega$. When  $k_{\rm B}T\gg\hbar\omega$, we recover the Johnson-Nyquist noise
\begin{equation}
\label{Johnson-Nyquist} S^{0}(x,\omega)|_{\omega\rightarrow
0}=2k_{\rm B}T\,G_{0},
\end{equation}
where we used that for small $\omega$ it holds that
$\coth(\beta\omega/2)=2/\beta\omega$. We also restored the units
of $\hbar$, i.e. $G_{0}=4e^2/h$. In the limit of $k_{\rm
B}T\ll\hbar\omega$ we obtain the quantum noise in the absence of
the scatterers
\begin{equation}
S^{0}(x,\omega)=\hbar|\omega|G_{0}.
\end{equation}
The full frequency dependence of $\sigma_{0}(x,x;\omega)$ contains
interference effects of the multiple plasmon reflection inside the
nanotube as well as interference terms depending on the measuring
point $x$. Instead of elaborating the noise of the clean system
further we concentrate on the impurity or shot noise, to be
discussed next \cite{noteDolcini}.

\subsubsection{Shot noise}
At zero frequency $\omega$, the impurity noise $S_{I}$  adds to
the total noise to give

\begin{equation}
\label{zeronoise} S=-e\,\coth\left(\frac{eV}{2k_{\rm
B}T}\right)\,I_{B}+2k_{\rm B}TG_{B}+2k_{\rm B}TG,
\end{equation}

where $G=G_{0}+G_{B}$ with $G_{B}=dI_{B}/dV$.
 At zero temperature, the shot noise
becomes $S=-e\coth(eV/2k_{\rm B}T)I_{B}$ with $\coth(eV/2k_{\rm
B}T)={\rm sgn}(eV)$ at $T=0$. We then finally obtain for the zero
temperature noise at zero frequency
\begin{equation}
S=e|I_{B}|.
\end{equation}
Note that it is the electron charge $e$ rather than the fractional
charge $ge$ in front of $I_{B}$ in contrast to the infinite SWNT
with an impurity \cite{Smithanoise}.
\subsubsection{Fano factor}
 Here we discuss the
experimentally relevant Fano factor $F=S/eI$ which is the ratio of
the noise to the full shot noise $eI$. The Fano factor is only
well defined for the shot noise part of  Eq.~(\ref{zeronoise})
which is $S-2k_{\rm B}T\,G$. This Fano factor can be written in
dimensionless quantities as
\begin{equation}
\label{Fanofactor} F=\frac{-\coth\left(\frac{eV}{2k_{\rm
B}T}\right)\frac{I_{B}}{I_{0}}+\frac{2k_{\rm
B}T}{eV}\frac{G_{B}}{G_{0}}}{\left(1+\frac{I_{B}}{I_{0}}\right)},
\end{equation}
where we used that the total current $I=I_{0}+I_{B}$. Note to be
consistent with the lowest order expansion in the backscattering,
we would have to expand the denominator in  Eq.~(\ref{Fanofactor})
and keep only $I_{0}$. However, this distinction is only essential
if the next order would contribute significantly.
\begin{figure}[h]
\begin{center}
\hbox{\resizebox{4.2cm}{!}{\includegraphics{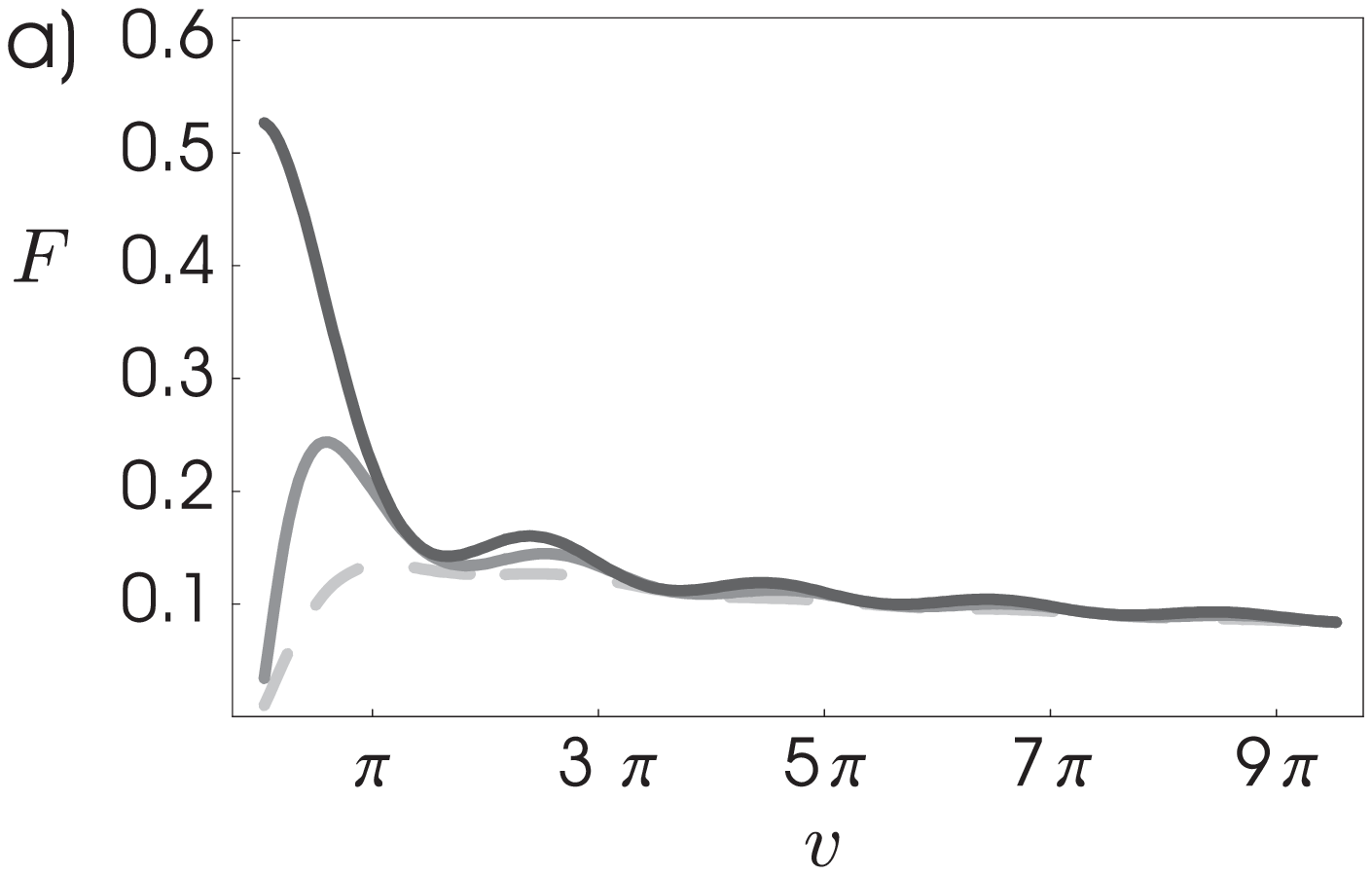}}\,\,\resizebox{4.35cm}{!}{\includegraphics{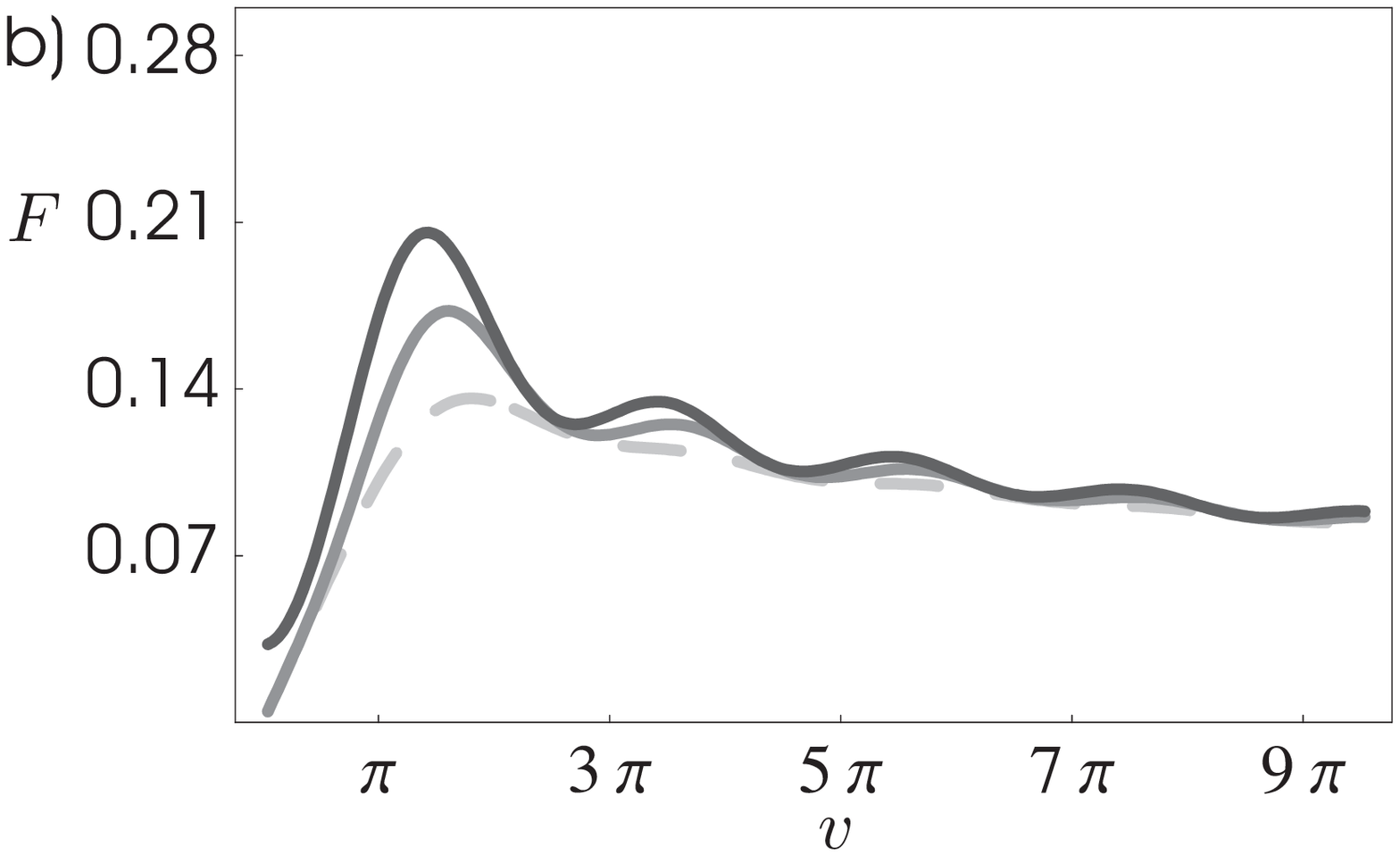}}}
\end{center}
\begin{center}
\hbox{\resizebox{4.2cm}{!}{\includegraphics{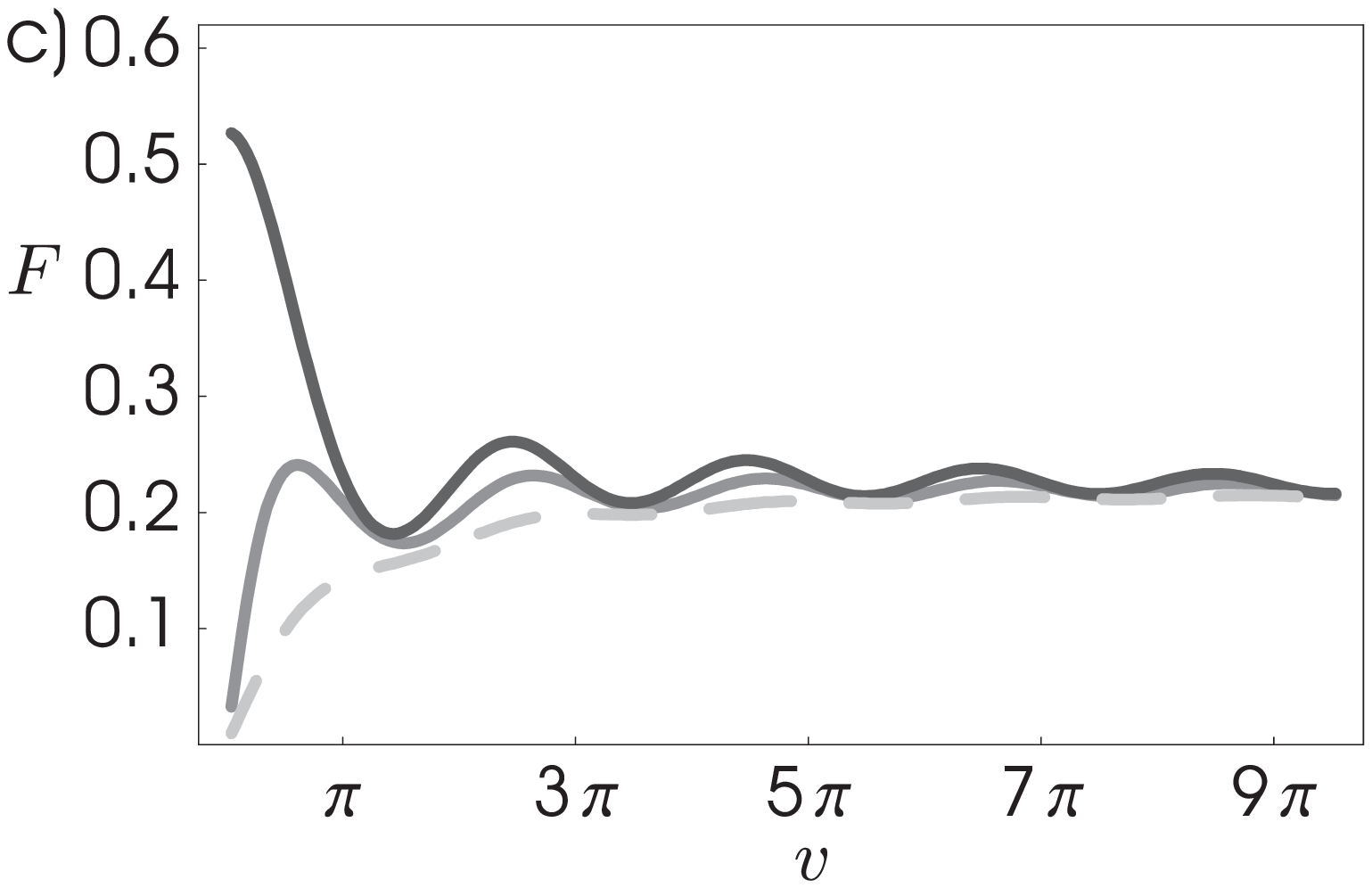}}\,\,\resizebox{4.35cm}{!}{\includegraphics{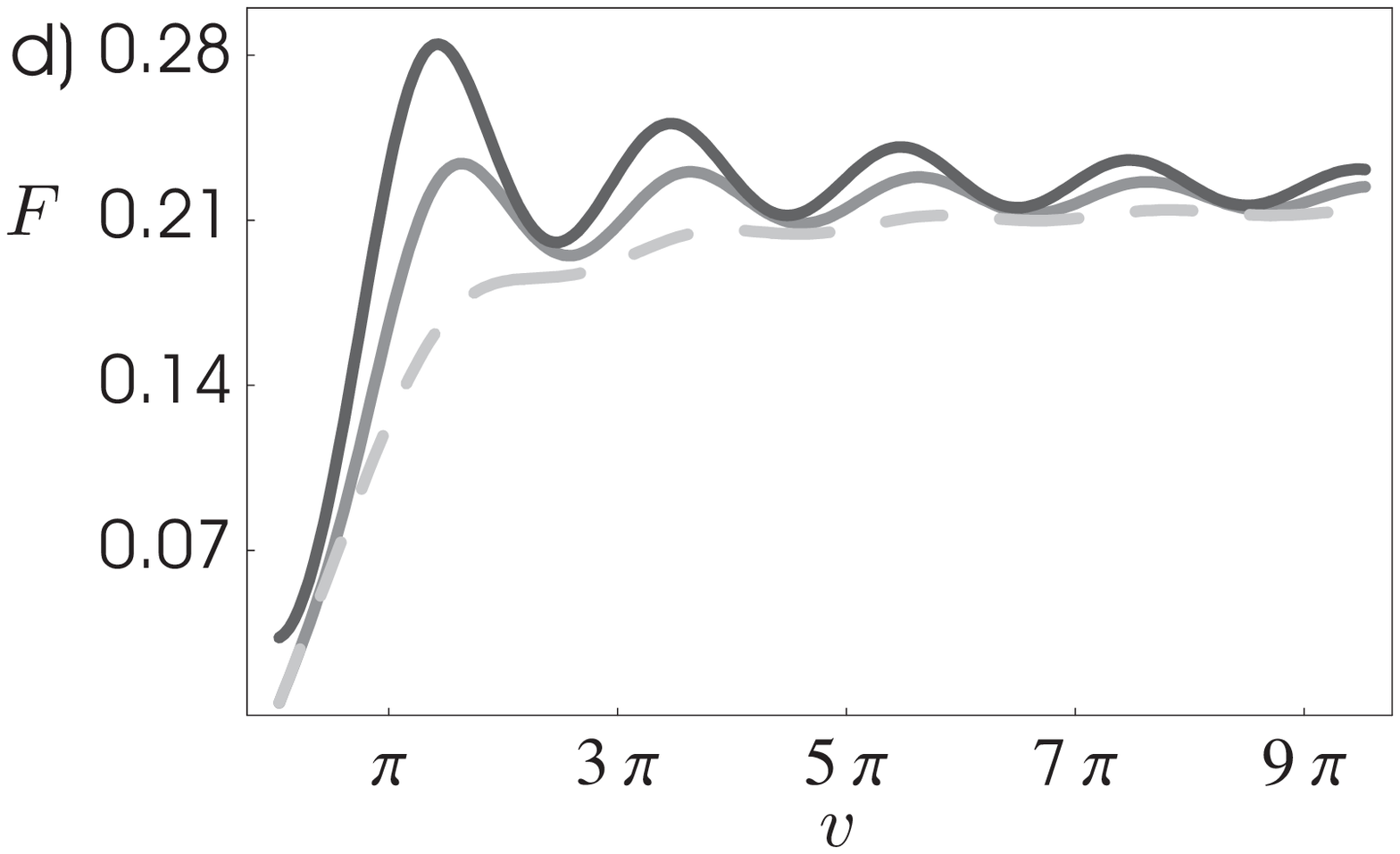}}}
\end{center}
\caption{The Fano factor $F$ defined in Eq.~(\ref{Fanofactor}) for
different temperatures and backscattering coefficients. In a) and
b)  we show the strongly correlated case with $g=0.23$ whereas in
c) and d) we present the non-interacting case $g=1$. In a) and c)
we use $U^{\rm in}=0.12$ and $U^{\rm co}=0.1$, in b) and d) we use
$U^{\rm in}=0.12$ and $U^{\rm co}=-0.1$ at the lowest temperature.
The temperatures in units of the non-interacting level spacing
$\hbar v_{\text{\textsc{f}}}/L$ are: $\Xi=0$(black),
$\Xi=0.3$(light gray) and $\Xi=0.7$(dashed).}
\end{figure}
\subsection{Analytical results of high-frequency noise}
Although an analytical solution of the time integrals in
Eq.~(\ref{sw2main}) is not possible we can estimate the general
trend for the impurity noise at high energies. We assume that the
temperature is close to zero, i.e. $|eV|,|eV\pm\hbar\omega|\gg
\hbar/t_{c},k_{\rm B}T$, and obtain for the incoherent part
\begin{widetext}
\begin{multline}
\label{analytichighfrequency} S_{I}^{\rm
in}=\sum\limits_{m=1,2}U_{m}^{\rm
in}\,|\sigma_{0}(x,x_{m};\omega)|^2\,\frac{\pi}{4}\sum\limits_{r=\pm}\coth\left(\frac{|v+r{\tilde
\omega}|}{2
\Xi}\right)\frac{|v+r{\tilde \omega}|^{1-\gamma/2}}{\Gamma(2-\gamma/2)}\\
\\
-2\sum\limits_{m}U_{m}^{\rm in}\coth\left(\frac{{\tilde
\omega}}{2\Xi}\right){\rm Re}\,\sigma_{0}(x,x_{m};\omega)
\frac{(\pi/4)}{\Gamma(2-\gamma/2)}\biggl\{{\rm
Re}\,\sigma_{0}(x,x_{m};\omega)\left({\rm sgn}({\tilde
\omega}-v)|{\tilde \omega}-v|^{1-\gamma/2}+{\rm
sgn}({\tilde \omega}+v)|{\tilde \omega}+v|^{1-\gamma/2}\right)\\
-{\rm Im}\,
\sigma_{0}(x,x_{m};\omega)\cot[\pi(2-\gamma)/4](|v+{\tilde
\omega}|^{1-\gamma/2}+|v-{\tilde
\omega}|^{1-\gamma/2}-2|v|^{1-\gamma/2})\biggr\}.\\
\end{multline}
\end{widetext}
\subsubsection{Analytical result for $g=1$}
It is worth to examine the non-interacting limit  $g=1$ in the
above expression where the asymptotic approximation
Eq.~(\ref{analytichighfrequency}) is exact (note that the
temperature dependence of the correlation functions does not
contribute when $g=1$, see Eq.~(\ref{fullT})). We show now that we
essentially get the Landauer-B\"uttiker result in this case,
namely

\begin{multline}
\label{non-interactingS}
 S_{I}^{{\rm in},g=1}=\frac{e^2}{2
h} \sum\limits_{n}R_{n}
\sum\limits_{r=\pm}\coth\left(\frac{eV+r\hbar\omega}{2k_{\rm
B}T}\right)
(eV+r\hbar\omega)\\
-\frac{2e^2}{h}\hbar\omega\coth\left(\frac{\hbar\omega}{2k_{\rm
B}T}\right)
\left\{\sum\limits_{n}R_{n}^{(1)}\cos^{2}[\omega(L+\Delta
x)/v_{\text{\textsc{f}}}]\right.\\
\left.+\sum\limits_{n}R_{n}^{(2)}\cos^{2}[\omega\Delta
x/v_{\text{\textsc{f}}}]\right\},
\end{multline}
where $\Delta x=x-L/2\geq 0$. In Eq.~(\ref{non-interactingS}) we
have written the noise in dimensionfull units, and we also
introduced reflection coefficients $R_{n}=R_{n}^{(1)}+R_{n}^{(2)}$
for barrier (1) and (2), respectively. They are related to $U^{\rm
in}$ by $\frac{e^2}{h}\sum_{n}R_{n}=(\pi/2)U^{\rm in }G_{0}$. This
choice is motivated by the fact that $I_{B}=-(1/2)I_{0}\pi U^{\rm
in}$ in the non-interacting limit (only incoherent contribution
considered). To make contact with the Landauer-B\"uttiker
formalism \cite{BuBla} we write the total current as
$I=(e^2/h)\sum_{n}T_{n} V$ with $T_{n}=1-R_{n}$ being the
transmission coefficient for mode $n$. In our regime of small
reflections ($R_{n}\ll 1$) we have $I=I_{0}+I_{B}$ and therefore
$I_{B}^{{\rm in},g=1}= -(e^2/h) \sum_{n}R_{n} V$. The result
Eq.~(\ref{non-interactingS}) agrees exactly with Eq.~(12) in
Ref.~40 in the limit of zero temperature and weak backscattering.
Eq.~(\ref{non-interactingS})  coincides with the
Landauer-B\"uttiker formalism only up to the oscillatory terms
which depend on the measurement position $x$ and become important
if $2\omega(L+\Delta x)/v_{\text{\textsc{f}}}\gtrsim 1$. This
oscillatory behavior for large frequencies, or equivalently, for
measurement points far away from the impurities results from the
beat note of finite frequency noise for energies which differ by
$\pm \hbar\omega$. The phase difference acquired from the
measurement point to the impurities and back will result in the
observed interference oscillations.

For $g=1$, we can calculate the interference contribution
proportional to  $U^{\rm co}$ in closed form. Using $(\pi/2)
U^{\rm co}G_{0}=2(e^2/h)\sum_{n}\sqrt{R_{n}^{1}R_{n}^{2}}$ and
dimensionfull units we obtain
\begin{widetext}
\begin{multline}
\label{nonintnoise} S_{I}^{{\rm co},g=1}=-\frac{e^2}{h}\frac{2\pi
k_{\rm B}T}{\sinh(2\pi k_{\rm B}T
t_{\text{\textsc{f}}}/\hbar)}\coth\left(\frac{\hbar\omega}{2k_{\rm
B}T}\right)
\sum\limits_{n}\sqrt{R_{n}^{1}R_{n}^{2}}\sum\limits_{m\neq
m'}\sum\limits_{r=\pm}{\rm Re}\,\sigma_{0}(x,x_{m};\omega)\\
\times\left\{{\rm Im}\,
\sigma_{0}(x,x_{m'};\omega)\cos\left[\frac{(\hbar\omega+reV)t_{\text{\textsc{f}}}}{\hbar}\right]
+{\rm Re}\,
\sigma_{0}(x,x_{m'};\omega)\sin\left[\frac{(\hbar\omega+reV)t_{\text{\textsc{f}}}}{\hbar}\right]-{\rm
Im}\,\sigma_{0}(x,x_{m};\omega)\cos\left(\frac{eVt_{\text{\textsc{f}}}}{\hbar}\right)\right\}\\
\\
 +\frac{e^2}{h}
\sum\limits_{n}\sqrt{R_{n}^{1}R_{n}^{2}}\,{\rm
Re}\left\{\sigma_{0}(x,x_{1};\omega)^{*}\sigma_{0}(x,x_{2};\omega)\right\}\sum\limits_{r=\pm}\coth\left(\frac{eV+r\hbar\omega}{2k_{\rm
B}T}\right)\frac{2\pi k_{\rm B}T}{\sinh(2\pi
k_{\rm B}Tt_{\text{\textsc{f}}}/\hbar)}\sin\left[\frac{(eV+r\hbar\omega)t_{\text{\textsc{f}}}}{\hbar}\right].\\
\end{multline}
\end{widetext}

\subsection{Physical interpretation of shot noise results}
We first discuss the results for the low-frequency noise. The
general result for the low-frequency noise is presented in
Eq.~(\ref{zeronoise}). This result, valid for finite temperatures,
is formally identical to that of an {\it infinite} TLL
\cite{KaneFisher94} except for the important difference that a
renormalization of the backscattered charge is absent. This fact
is not at all trivial and has first been predicted by Ponomarenko
and Nagaosa \cite{Nagaosa} in the case of a random backscattering
potential. Our result shows that this is also true for the SWNT
with double barriers. One could argue that the high
bias/temperature transport regime is sensitive to interaction (see
subsection IV B) and thus a charge $g_{{\rm eff}}e$ in front of
the backscattered current appears, rather than the charge $e$.
However, this conclusion is wrong. Even if a fractional charge is
locally backscattered by the barriers, this charge cannot directly
enter the leads, but will further get partially backscattered at
the interfaces due to the inhomogeneity of the TLL parameter $g$.
Summing over all backscattered partial charges results in the
electron charge $e$. The zero-frequency noise is only sensitive to
this integral effect as it sums up correlations over all times.
Therefore, at low frequencies, the transport process is that of
electrons with charge $e$ which are backscattered by a scattering
region connecting two Fermi liquid leads. It is interesting that
this conclusion is independent of the bias voltage; the regime of
low or high bias voltage is only distinguished by the power-laws
of transmission. The most pronounced  effect of interactions in
low-frequency noise or Fano factor is its power-law dependence on
bias voltage and temperature.

\begin{figure}[h]
\label{highfrequency1}
\begin{center}
\includegraphics[width=5.5cm]{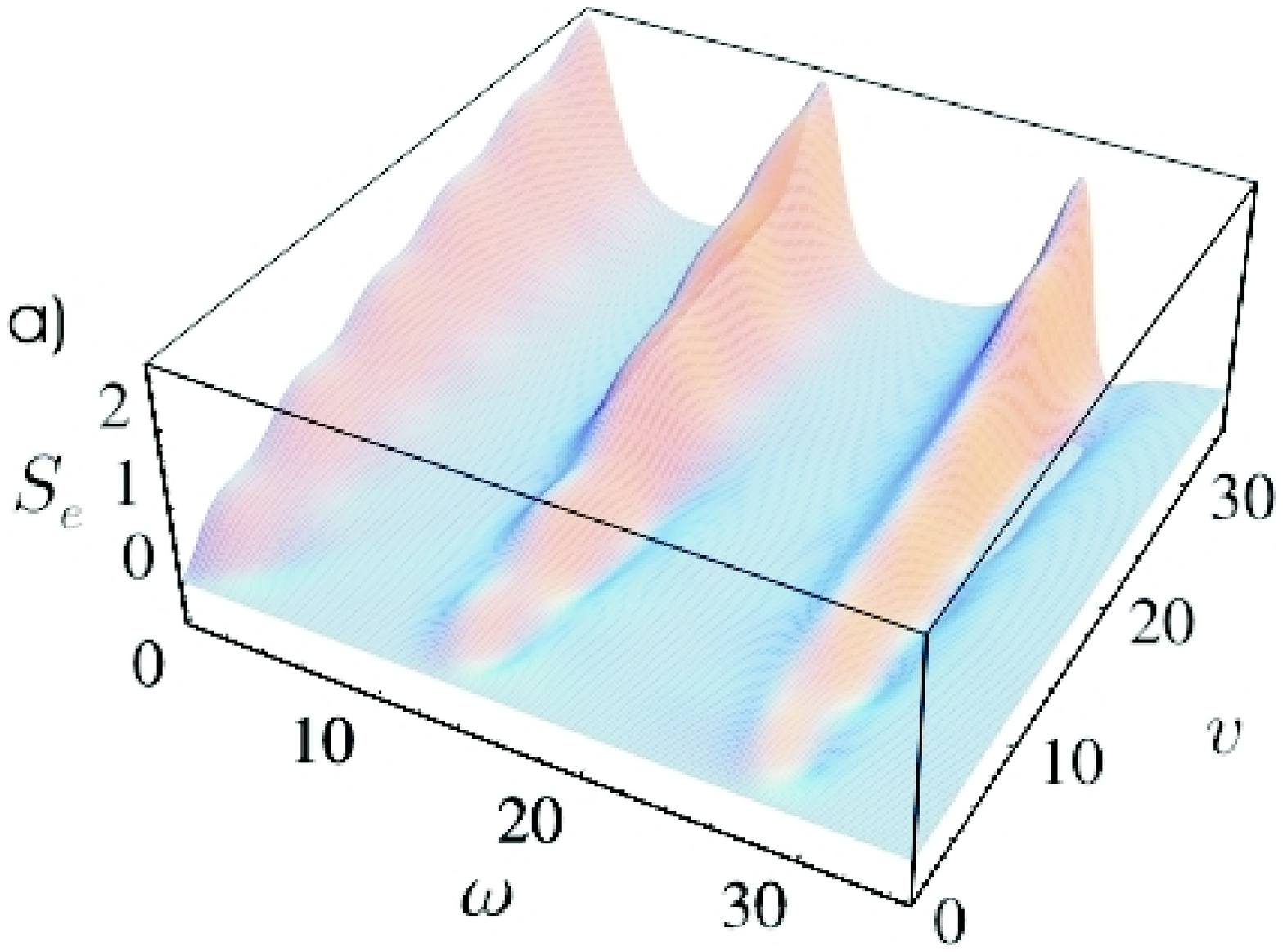}
\end{center}
\begin{center}
\hbox{\resizebox{4.2cm}{!}{\includegraphics{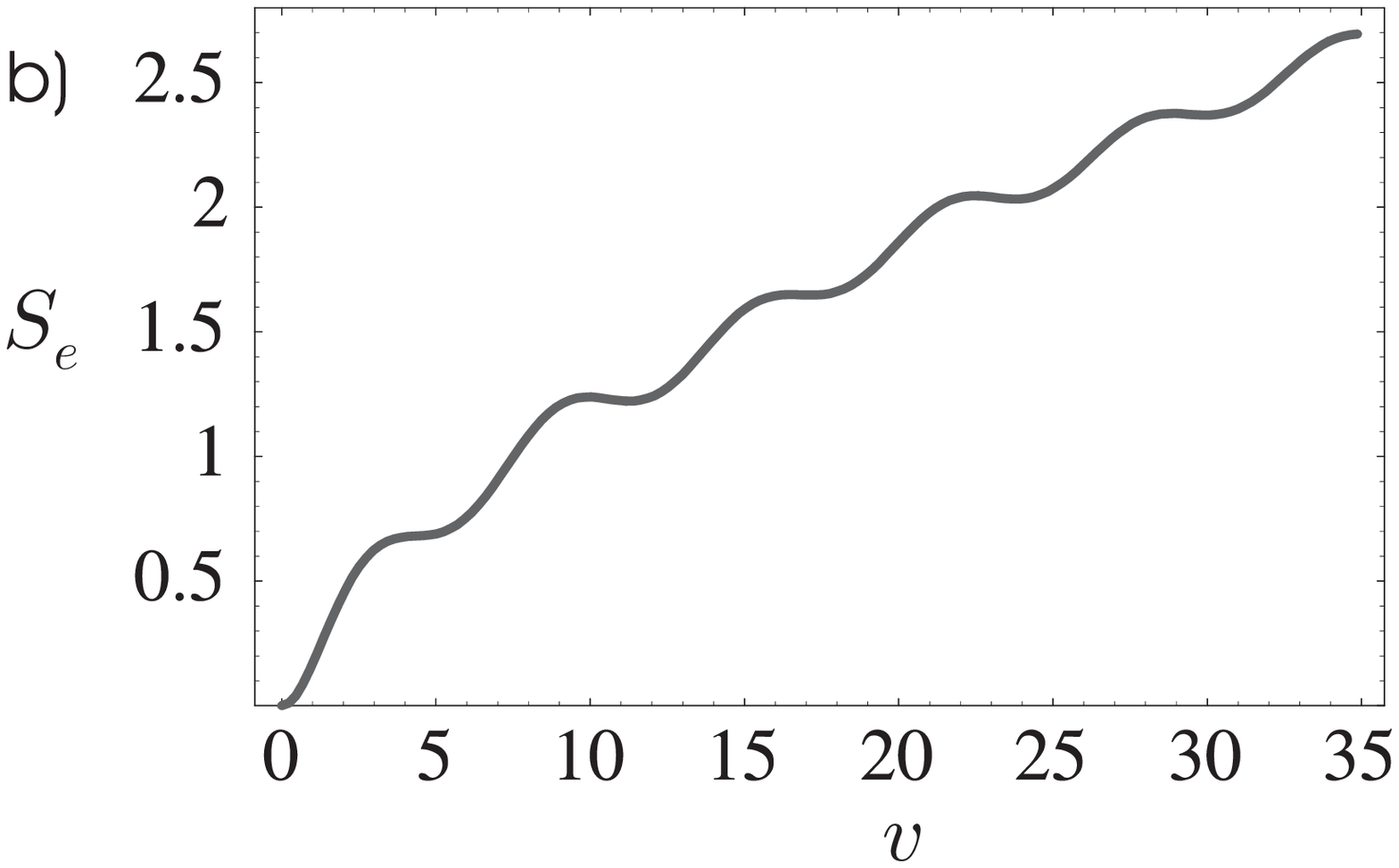}}\quad\resizebox{4.2cm}{!}{\includegraphics{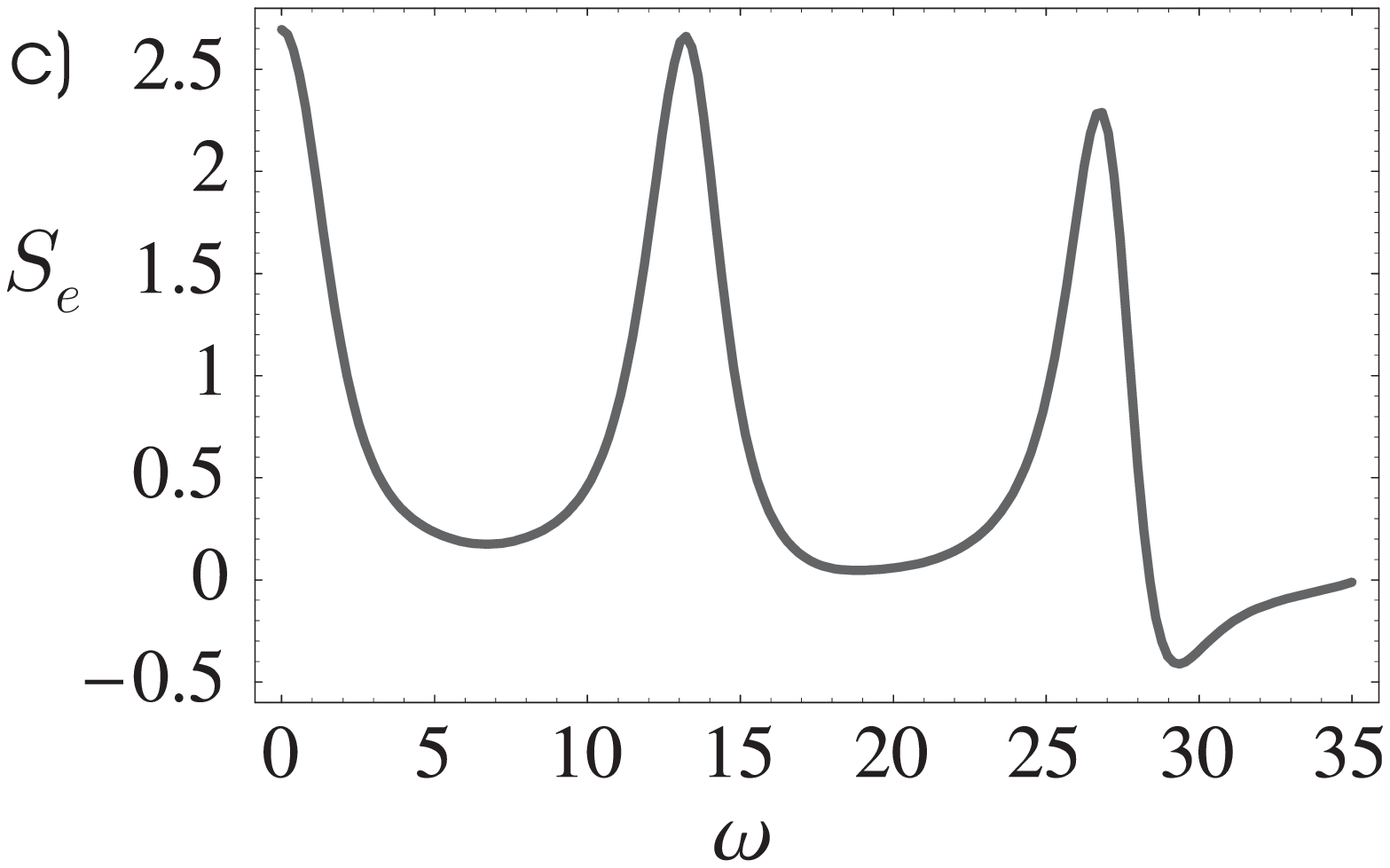}}}
\end{center}
\caption{(color online). Excess noise
$S_{e}=S(V,\omega)-S(0,\omega)$ in units of
$G_{0}\hbar/t_{\text{\textsc{f}}}$ at temperature $\Xi=0.3$ for
interaction strength $g=0.23$ using numerical integration of
Eq.~(\ref{sw2main}). The backscattering coefficients are
$U_{1}^{\rm in}=U_{2}^{\rm in}=0.06$ and $U^{\rm co}=0.1$. We have
chosen the measurement point to be at one of the barriers. Figure
a) shows the excess noise as a function of bias voltage $eV$ and
frequency $\hbar\omega$  (both in units of the non-interacting
level spacing $\hbar v_{\text{\textsc{f}}}/L$). In Plot b) we
present the low frequency noise $\omega=0$ as a function of bias
voltage. Clear FP-oscillations dominated by the non-intearcting
frequency $v_{\text{\textsc{f}}}/L$ are seen as well as the
power-law scaling with power $1-\gamma/2$. In Plot c) we show the
frequency-dependence of excess noise at large bias voltage $v\sim
34$. Dominant charge mode oscillations with frequency $\hbar
v_{\text{\textsc{f}}}/2Lg$ are observed.}
\end{figure}
\begin{figure}[h]
\label{highfrequency2}
\begin{center}
\includegraphics[width=5.5cm]{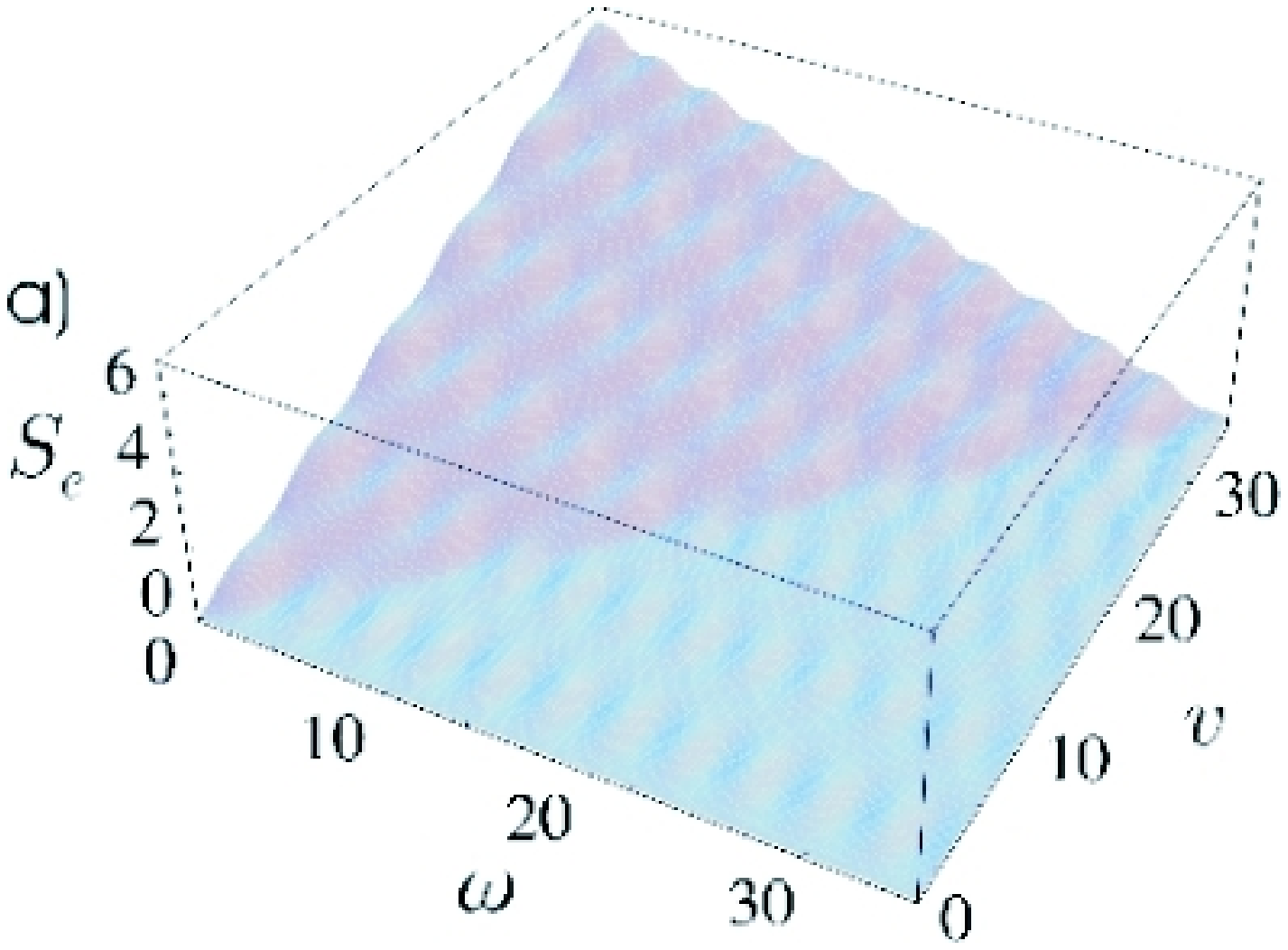}
\end{center}
\begin{center}
\hbox{\resizebox{4.2cm}{!}{\includegraphics{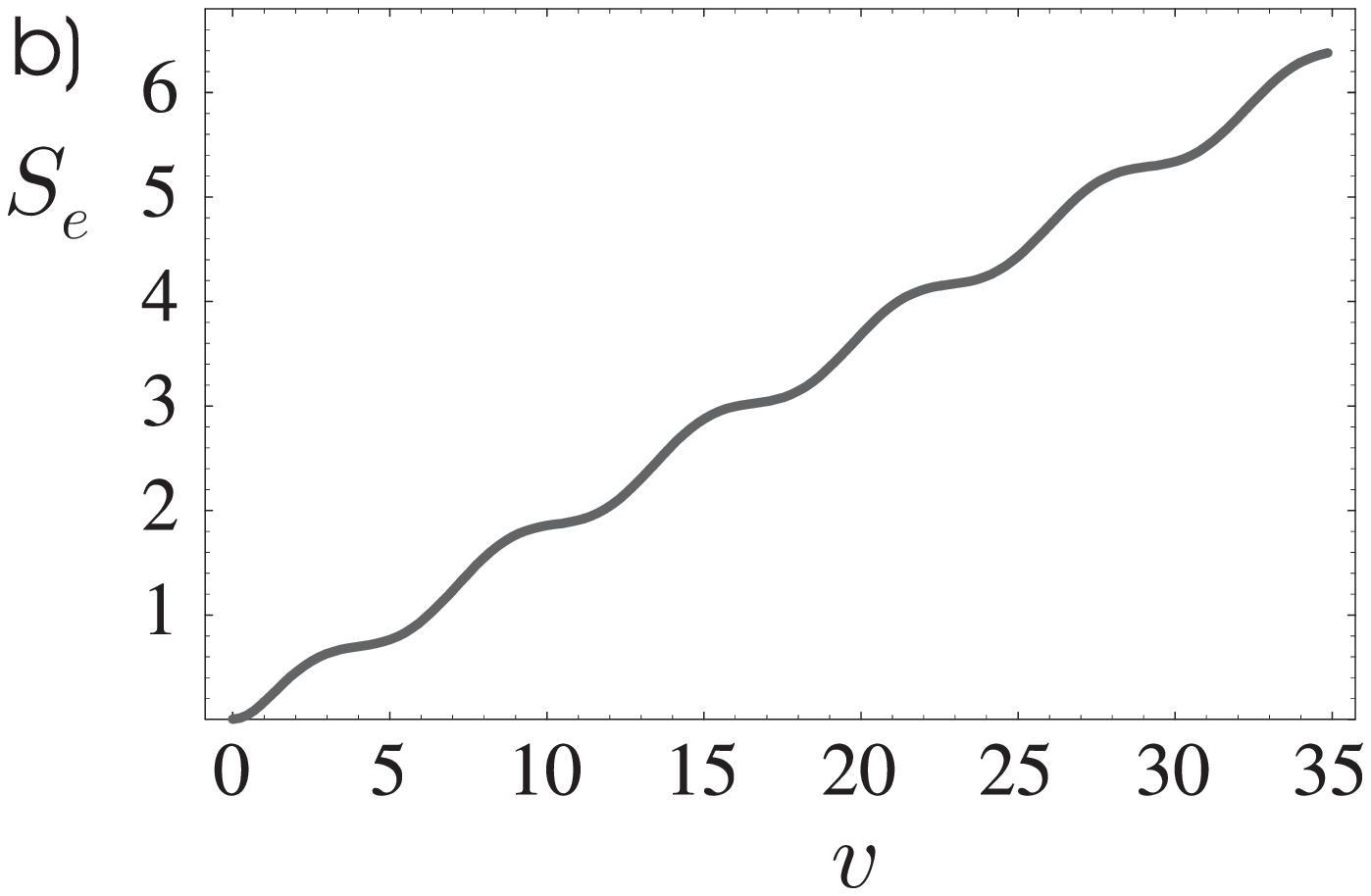}}\quad\resizebox{4.2cm}{!}{\includegraphics{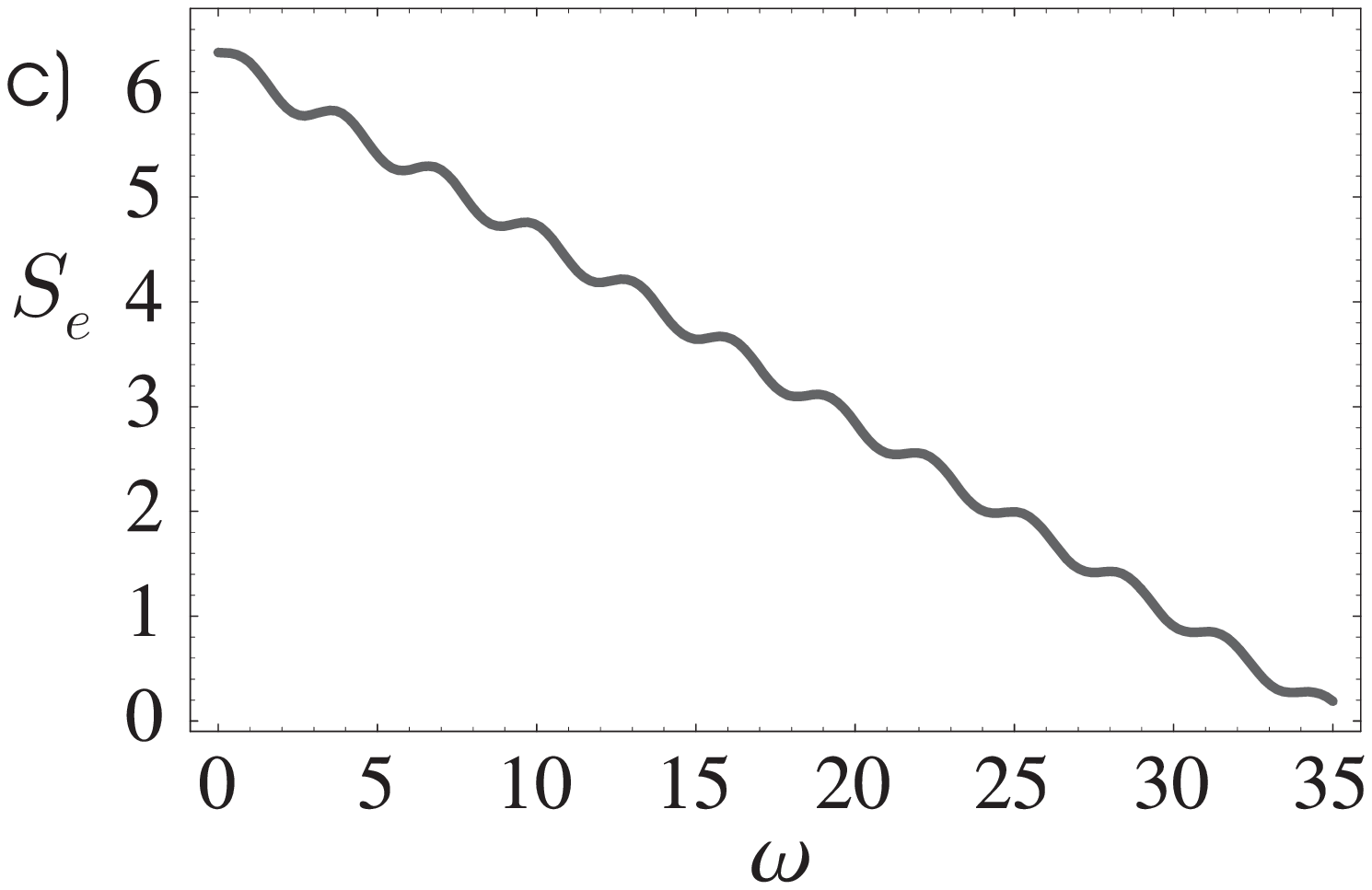}}}
\end{center}
\caption{(color online). Same parameters as in Fig.~7 but for a
non-interacting system with $g=1$. In graph b) approximately the
same periodicity of FP-oscillations is found as in the interacting
case but the frequency dependence at large bias voltage presented
in graph c) is clearly different. Only much smaller ordinary
FP-oscillations with frequency $v_{\text{\textsc{f}}}/L$ are seen.
Superimposed are the additional oscillations depending on the
measurement point. For the chosen point of measurement $|x|=L/2$
this oscillation frequency is $v_{\text{\textsc{f}}}/2L$. }
\end{figure}
The finite frequency impurity noise Eq.~(\ref{sw2main}) contains
FP-oscillations coming from all collective modes as well as a
periodic noise suppression as a function of frequency with the
oscillation period determined by the charge mode velocity $v_{c}$.
At large bias voltage, this frequency dependence is dominated by
the first term on the right-hand-side of
Eq.~(\ref{analytichighfrequency}). The periodic modulation
originates from $|\sigma_{0}(x,x_{m};\omega)|^2$ where $x$ is the
point of measurement. Interestingly, this term does not depend on
$x$ in contrast to the terms $\propto \coth({\tilde \omega}/2\Xi)$
in Eq.~(\ref{analytichighfrequency}), which, however, are smaller
when $eV\gg \hbar\omega$.
\begin{figure}[h]
\begin{center}
\includegraphics[width=5.9cm]{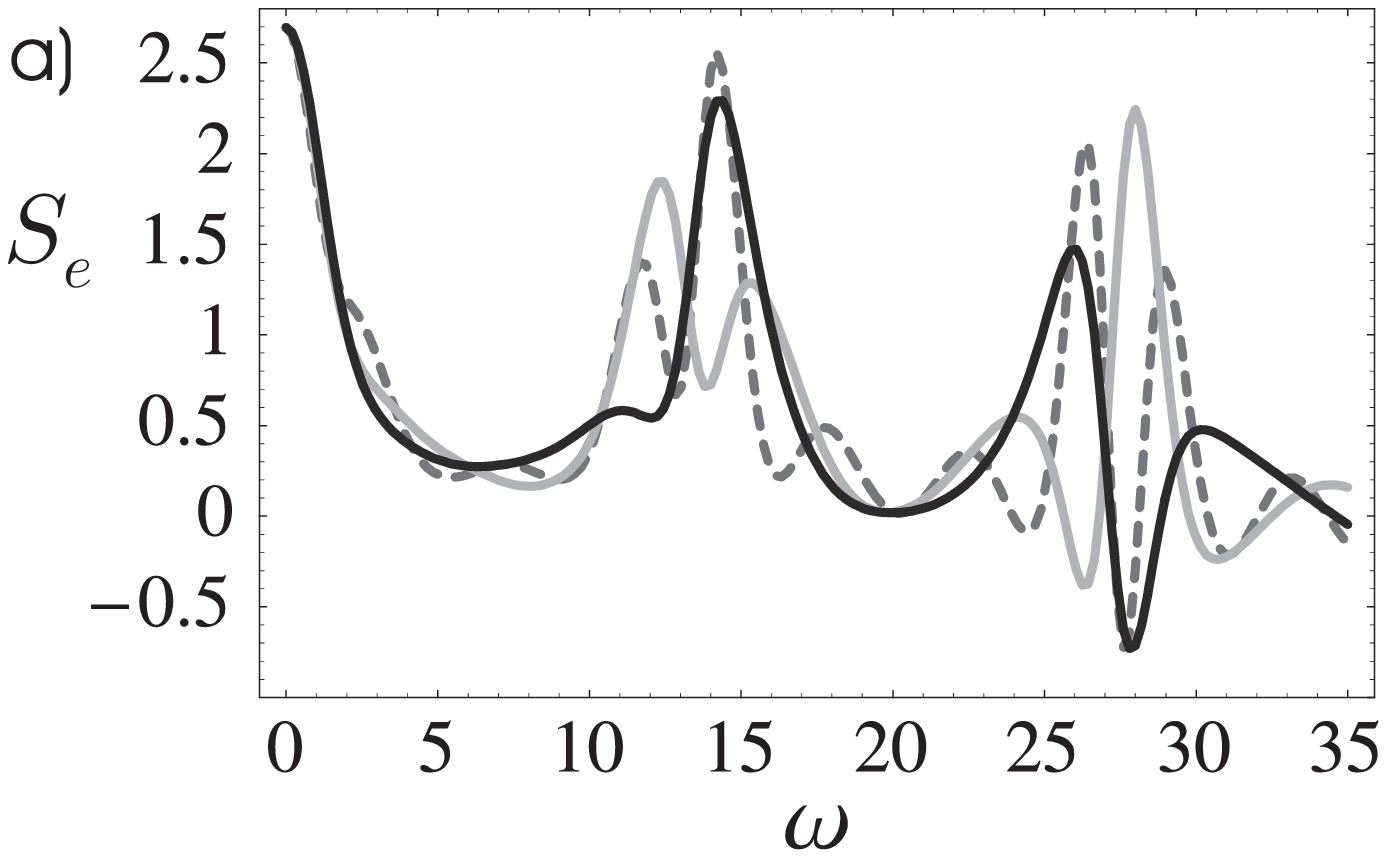}
\end{center}
\begin{center}
\includegraphics[width=5.7cm]{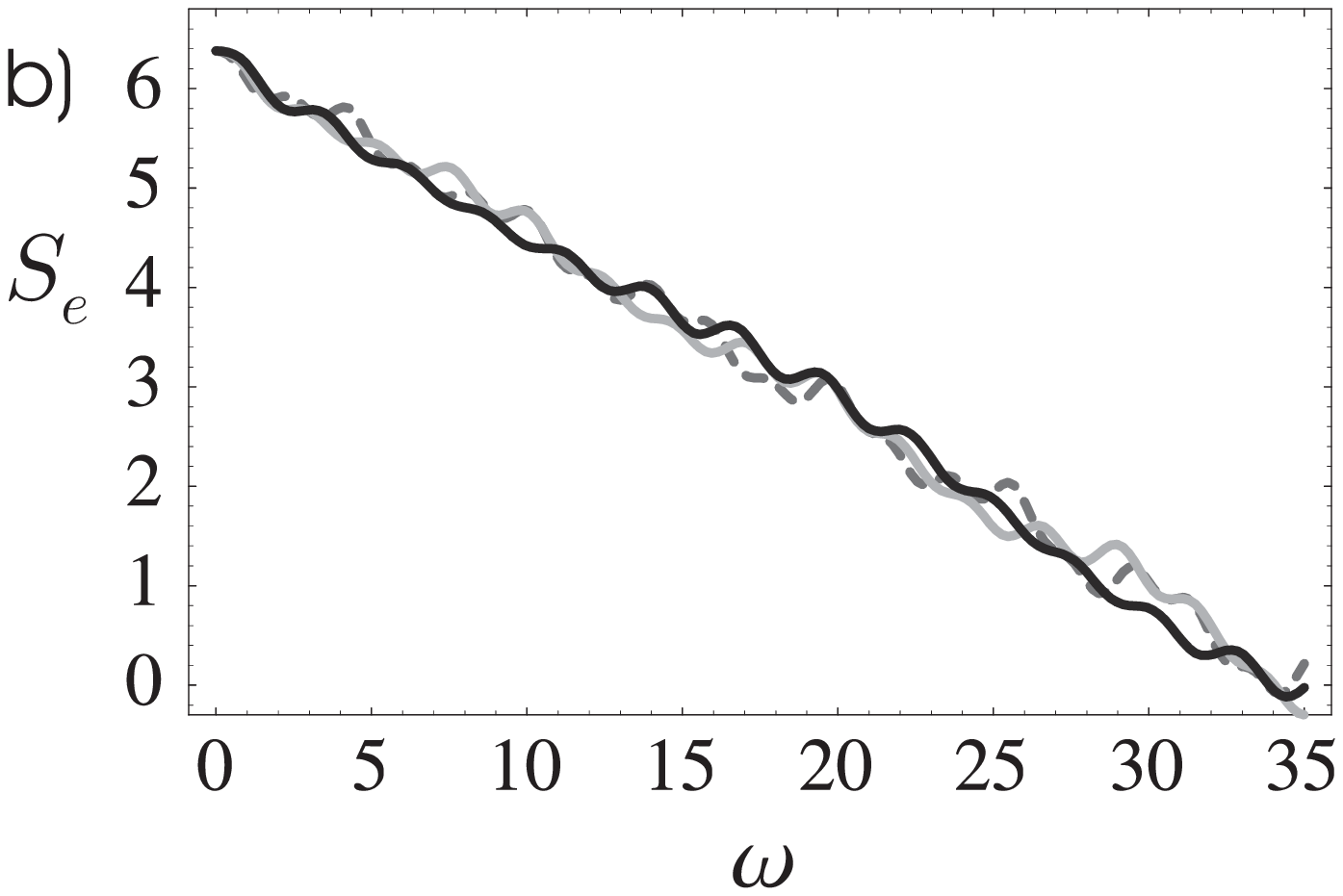}
\end{center}
\caption{The excess noise $S_{e}$ in the same regime as in
Figs.~7c) and 8c) but for different measurement points in the
leads $|x|> L/2$. In graph a) we present the strongly interacting
case $g=0.23$ and in graph b) the non-interacting case $g=1$. We
show curves for three different measurement points:
$d=0.14$(dark), $d=0.3$(light gray) and $d=0.6$(dashed) where $d$
is the distance from $x$ to the nearest barrier in units of
SWNT-length $L$.}
\end{figure}
Therefore, at high bias voltage and/or low frequencies, the noise
clearly shows the oscillations with frequency $v_{c}/2L$. This is
indeed observed in Fig.~7. These oscillations are a consequence of
the charge fragmentation at the SWNT-metal reservoir interfaces
due to the inhomogeneity of $g$. These oscillations have to be
distinguished from the oscillations due to standard
FP-interferences which contain the frequencies of all modes. At
larger frequency $\omega$, the terms $\propto \coth({\tilde
\omega}/2\Xi)$ in Eq.~(\ref{analytichighfrequency}) become
important as well. They contain shot noise parts as well as
thermal noise parts. This is clear when noting that at zero
frequency these terms are the impurity dependent parts of $2k_{\rm
B}TG_{B}+2k_{\rm B}TG$ in Eq.~(\ref{zeronoise}). With growing
frequencies $\omega$, these terms become sensitive to the
measurement point $x$. This is clearly seen in the solution for
$g=1$ presented in Eqs.~(39) and (40). These oscillating factors
are a consequence of interference of electron waves which differ
in energy by $\hbar\omega$. On the way from the measuring point to
the impurities and back these waves will pick up different phases
which then results in the oscillation factors, see also Refs.~39
and 23. In principle, these oscillations will influence the noise
at every frequency, provided the measurement is taken far away
from the scattering region. In reality, however, the metal
contacts are not ballistic and therefore these results are only
valid for a measurement point $x$ near the barriers (within the
inelastic mean-free path of the contacts)\cite{Dolcini}.

We further comment on  Figs~7 and 8 which show the excess noise
$S_{e}(V,\omega)=S(V,\omega)-S(0,\omega)$ at low temperature
chosen to be $\Xi=0.3$, relevant for experiments. This definition
subtracts the noise in the absence of the barriers.  In the
non-interacting case (Fig. 8) the most striking features are the
clear diagonal structure in the 3D-plot 8a)  which states that the
excess noise is essentially zero when $\hbar\omega>eV$, see
Eq.~(\ref{non-interactingS}). The small oscillations in the excess
noise originate  from the interference term
Eq.~(\ref{nonintnoise}) and contain the FP-oscillations in both,
bias voltage  and frequency as well as oscillations as a function
of frequency depending on the measurement point [see dependence on
$d$ in Fig. 9b)]. The multiple reflections at the boundary between
nanotube and contacts are driven by the e-e interactions and
therefore these oscillations are absent when $g=1$. Fig.~7 shows
the excess noise for a strongly correlated system with $g=0.23$.
The low-frequency noise as a function of bias voltage [Fig.~7b)]
shows a power-law behavior with exponent $1-\gamma/2$ as well as
some minor qualitative differences of the  FP-interference
oscillations compared to the non-interacting case. But the
oscillation period is dominated by the non-interacting frequency.
The bias window $\Delta V$ between two maxima is very well
approximated by $\Delta V=h/t_{\text{\textsc{f}}}e$, i.e. the
voltage difference expected for a non-interacting system. The
frequency dependent noise at high bias voltage is clearly
different from the non-interacting case. Striking are the
oscillations of noise with period $\Delta\omega=\pi/t_{c}$ due to
the charge flux fragmentation at the SWNT-metal reservoir
interfaces. They are much more pronounced than the ordinary
FP-oscillations due to two barriers if $eV\gg \hbar
v_{\text{\textsc{f}}}/L$ since $I_{B}^{\rm in}$ grows
monotonically with bias voltage whereas the strength of
$I_{B}^{\rm co}$ (showing the FP-oscillations) is bounded roughly
by the level spacing $\hbar v_{\text{\textsc{f}}}/L$. At low
frequencies, these oscillations are clearly resolved. At larger
frequencies, oscillations depending on the measurement point are
superimposed [see Fig. 9a)]. However, the charge mode oscillation
period $\pi/t_{c}$ is still very pronounced. Note, the excess
noise is not zero anymore at large frequencies as this is the case
in the absence of interaction. This is mainly due to the last term
in the asymptotic formula Eq.~(\ref{analytichighfrequency}) which
depends on the bias voltage. Therefore, the excess noise receives
a non-zero contribution from this term. Note that its prefactor
$\cot[\pi(2-\gamma)/4]$ vanishes for $g=1$, and, as a consequence,
this contribution is absent when $g=1$. The excess noise can even
get negative in agreement with Ref.~23. Even if $S_{e}$ is not
strictly zero for $\hbar\omega>eV$, there is still a pronounced
diagonal structure showing a cusp singularity when
$eV=\hbar\omega$.

 To observe the high-frequency oscillations we
must at least be able to see the first minimum, which translates
into $\Delta\omega\sim \pi/2t_{\text{\textsc{f}}}g$. To have still
ballistic transport we should be well below the mean free path of
a carbon nanotube which, at low temperatures can exceed several
micrometers. Using $L\sim 10\mu{\rm m}$ this translates into an
estimate for the frequency of $\omega/2\pi\lesssim$ 100 GHz which
is in the range of existing technology \cite{Schoelkopf,Deblock}.
A relevant extension of this setup over previously discussed
systems \cite{Dolcini} is achieved through the inclusion of two
impurities as well as through the consideration of four modes
relevant for a carbon nanotube. In this system, ordinary
FP-oscillations as well as oscillations in noise due to the finite
length of the interacting region {\it coexist}. This allows us to
extract the TLL parameter $g$ by comparing the oscillations in
bias voltage and as a function of frequency, i.e. building the
ratio $\Delta V/\Delta\omega\sim (2\hbar/e) g$ allows to estimate
$g$ without referring to power-law fitting and without knowing of
any other system parameter like the length $L$ or the precise
position of an impurity in the wire \cite{Dolcini}. We should also
mention, that this ratio $\Delta V/\Delta\omega$ contains valuable
information about spin-charge separation or in general,
information about different velocities of elementary excitations
in a carbon nanotube.

\section{Conclusions}
We have calculated and discussed in detail conductance and finite
frequency shot noise in a single-walled carbon nanotube (SWNT) in
good contact to electron reservoirs using a non-equilibrium
Keldysh functional integral approach. Special focus was put on the
interference of backscattering events off two weak impurities
naturally formed at the interface between the SWNT and metal
contacts. These so called Fabry-Perot (FP) interferences exhibit
oscillations in conductance and shot noise as a function of bias
voltage and noise frequency which are dominated by the
non-interacting traversal time
$t_{\text{\textsc{f}}}=L/v_{\text{\textsc{f}}}$ rather than the
interacting velocity $t_{c}=t_{\text{\textsc{f}}}g$, with $L$ the
SWNT-length and $g$ the Tomonaga-Luttinger liquid parameter, due
to two degenerate subbands in the SWNT. However, the finite
frequency noise is in addition capable to resolve the splintering
(momentum-conserving reflections of fractional charge) of the
transported electrons due to the finite length of the interacting
SWNT. This dynamics leads to oscillations in the frequency
dependent excess noise, which, at large bias voltages, are
dominated by a single frequency $(2t_{c})^{-1}$, despite the
existence of the ordinary FP-interference oscillations. Therefore,
shot noise measurements as a function of bias voltage and
frequency seem a decisive tool to distinguish the two mode
velocities in the SWNT.
\section{Acknowledgments}
P. Recher would like to thank L. Balents, M.P.A. Fisher, H.
Grabert and N. Nagaosa for helpful comments and discussions. This
work is supported by JST/SORST, NTT, University of Tokyo and
ARO-MURI grant DAAD19-99-1-0215.
\appendix
\section{Construction of the Keldysh action $S_{0}$}
The Keldysh-action $S_{0}$ introduced in
Eq.~(\ref{generfunctional}) is constructed in terms of equilibrium
correlation functions between the Keldysh fields. We need the
following correlation function (e.g. for the
$\theta\phi$-correlation)
\begin{eqnarray}
\label{correlation} C^{\theta\phi}_{a}(x,x';t)&=&\langle
{\hat T}_{K}\theta_{a}(x,t)\phi_{a}(x',0)\rangle_{0}\nonumber\\
&=&\frac{1}{2}\langle\{{\theta}_{a}(x,t),{\phi}_{a}(x',0)\}\rangle_{0},
\end{eqnarray}
and the retarded functions
\begin{eqnarray}
\label{response} R^{\theta\phi}_{a}(x,x';t)&=&\langle{\hat
T}_{K}\theta_{a}(x,t){\tilde
\phi}_{a}(x',0)\rangle_{0}\nonumber\\
&=&-i\Theta(t)\,\langle[{\theta}_{a}(x,t),{\phi}_{a}(x',0)]\rangle_{0}.
\end{eqnarray}

The expectation values are taken at equilibrium $V=0$ and in the
absence of backscattering $(H_{\rm bs}=0)$. Other combinations
like $\langle{\hat T}_{K}{\tilde \theta}_{a}(x,t){\tilde
\phi}_{a}(x',0)\rangle=\langle{\hat T}_{K}{\tilde
\phi}_{a}(x,t){\tilde \theta}_{a}(x',0)\rangle=0$. Since the
expectation values are determined by the dynamics of
$H_{\text{\textsc{SWNT}}}$ only,  different sectors $a=1...4$ do
not mix, i.e. $\langle{
\theta}_{a}(x,t){\phi}_{a'}(x',0)\rangle_{0}=0$ for $a\neq a'$.
The action $S_{0}$ then can be written as
\begin{widetext}
\begin{equation}
\label{action}
 S_{0}=\frac{i}{2}\sum\limits_{a=1}^{4}\int dx\int dx'\,\int
d\omega\,\left[\theta_{\bf a}^{\rm T}(x,-\omega),\phi_{\bf a}^{\rm
T}(x,-\omega)\right]\, G_{\bf
a}^{-1}(x,x';\omega)\left[\begin{array}{c}\theta_{\bf
a}(x',\omega)\\\phi_{\bf a}(x',\omega)\end{array}\right].
\end{equation}
In Eq.~(\ref{action}) the vector ${\bf \theta_{a}}(x,\omega)$ is
defined as $ \theta_{\bf
a}(x,\omega)=[\theta_{a}(x,\omega),{\tilde
\theta_{a}}(x,\omega)]^{\rm T}$, and similar for $\phi_{\bf
a}(x,\omega)$, where, here, T means the matrix transpose. The
matrix of the Green's function operator  $G_{\bf
a}(x,x';\omega)=\langle x|{\bf G_{ a}}(\omega)|x'\rangle$ is
constructed out of the equilibrium correlators and has the
representation
\begin{equation}
\label{Greenmatrix} G_{\bf
a}(x,x';\omega)=2\pi\,\left(\begin{array}{c}C_{a}^{\theta\theta}(x,x';\omega)\\R_{a}^{\theta\theta}(x',x;-\omega)\\C_{a}^{\phi\theta}(x,x';\omega)\\R_{a}^{\theta\phi}(x',x;-\omega)
\end{array}\begin{array}{c}R_{a}^{\theta\theta}(x,x';\omega)\\0\\R_{a}^{\phi\theta}(x,x';\omega)\\0\end{array}\begin{array}{c}C_{a}^{\theta\phi}(x,x';\omega)\\R_{a}^{\phi\theta}(x',x;-\omega)
\\C_{a}^{\phi\phi}(x,x';\omega)\\R_{a}^{\phi\phi}(x',x;-\omega)\end{array}\begin{array}{c}R_{a}^{\theta\phi}(x,x';\omega)\\0\\R_{a}^{\phi\phi}(x,x';\omega)\\0\end{array}\right).
\end{equation}
It obeys the symmetry  $G_{\bf a}(x,x';\omega)=G_{\bf a}^{\rm
T}(x',x;-\omega)$ which follows from the property
$C_{a}^{\theta\phi}(x,x';\omega)=C_{a}^{\phi\theta}(x',x;-\omega)$,
(and similar for $\theta\theta$ and $\phi\phi$-correlations) which
is evident from the defining  Eq.~(\ref{correlation}).

\section{Details of noise calculation}
In this appendix, we provide some details of the noise
calculations. Starting from Eq.~(\ref{noise1}) we perform the
functional derivatives and obtain to leading order in $H_{\rm bs}$

\begin{multline}
\label{fullomega}
S(x,\omega)=\frac{4}{\pi^2}e^2\omega^{2}C_{1}^{\theta\theta}(x,x;\omega)
-\frac{8}{\pi^2}e^2\,\sum\limits_{\bf n}\sum\limits_{mm'}\,\int
dt\,u_{m}^{\bf n}u_{m'}^{\bf n}\cos(eVt+\varphi_{mm'}^{{\bf n}})\\
\times\sum\limits_{rr'=\pm}rr'\,f_{m}^{r}(x,\omega)\left[f_{m}^{r}(x,-\omega)-e^{-i\omega
t}f_{m'}^{r'}(x,-\omega)\right]\,\left\langle e^{i[A_{m{\bf
n}}^{r}(t)-A_{m'{\bf n}}^{r'}(0)]}\right\rangle_{0},
\end{multline}
\end{widetext}
where
$f_{m}^{\pm}(x,\omega)=\omega\,[C_{1}^{\theta\theta}(x_{m},x;\omega)\pm\frac{i}{2}R_{1}^{\theta\theta}(x,x_{m};-\omega)]$
and the phase operators are
$A_{mijs}^{\pm}(t)={\theta}_{1m}^{\pm}+s\theta_{2m}^{\pm}+(-1)^{i+1}\delta_{ij}(\theta_{3m}^{\pm}+s\theta_{4m}^{\pm})
+(-1)^{i+1}(1-\delta_{ij})(\phi_{3m}^{\pm}+s\phi_{4m}^{\pm})$ and
$\varphi_{mm'}^{{\bf n}}=(V_{\rm g}L+2\Delta_{{\bf n}})(m-m')$.
Further, we introduced the index ${\bf n}=ijs$ and the
abbreviation $\langle ...\rangle_{0}=\prod_{a}\int{\cal
D}[\theta_{a}^{\pm}\phi_{a}^{\pm}]...\exp[iS_{0}]$.
 We assume now that the current is measured in the right
lead (the result for $x$ in the left lead is easily obtained by
$x_{1}\leftrightarrow x_{2}$ and $x\rightarrow -x$), i.e. $x\geq
L/2$. The general expression for the retarded function in that
case reads (see Appendix C)
\begin{multline}
\label{retardedfu}
R_{1}^{\theta\theta}(x,x_{m};\omega)=-\frac{i\pi}{2{
\omega}}(1-\gamma)\frac{e^{i\omega(\frac{x}{L}-\frac{1}{2})t_{\text{\textsc{f}}}}}{1-\gamma^2e^{i2\omega
t_{c}}}\\
\times\left(e^{i\omega(\frac{1}{2}-\frac{x_{m}}{L})t_{c}}+\gamma
e^{i\omega(\frac{3}{2}+\frac{x_{m}}{L})t_{c}}\right).
\end{multline}
Note that
$R_{1}^{\theta\theta}(x,x_{m};\omega)^{*}=R_{1}^{\theta\theta}(x,x_{m};-\omega)$.
The correlation functions are related to the retarded functions
via the fluctuation dissipation theorem
$C_{1}^{\theta\theta}(x,x_{m};\omega)=C_{1}^{\theta\theta}(x_{m},x;\omega)=-\coth(\beta\omega/2)\,{\rm
Im}R_{1}^{\theta\theta}(x,x_{m};\omega)$, where ${\rm Im}$ denotes
imaginary part. The expectation value in Eq.~(\ref{fullomega}) is
of the general form ($\varepsilon,\varepsilon'=+,-$)
\begin{equation}
\label{square} \left\langle e^{i\varepsilon A_{m{\bf
n}}^{r}(t')}e^{-i\varepsilon'A_{m'{\bf
n}}^{r'}(t'')}\right\rangle_{0}=e^{-\frac{1}{2}\langle[A_{m{\bf
n}}^{r}(t')-\varepsilon\varepsilon'A_{m'{\bf
n}}^{r'}(t'')]^{2}\rangle_{0}},
\end{equation}
where we used that the action $S_{0}$ is quadratic in the bosonic
fields $\theta_{a}^{\pm}$ and $\phi_{a}^{\pm}$ which allows to
perform the average in the exponent. The correlator in the
exponent of the right-hand-side (RHS) of Eq.~(\ref{square}) always
(i.e. for general ${\bf n}$) contains a sum of three
non-interacting modes $a=2,3,4$ due to spin and subband degeneracy
plus one interacting mode of the total charge $a=1$:
\begin{multline}
\label{f1}
 \left\langle\left(A_{m{\bf n}}^{r}(t')\pm A_{m'{\bf
n}}^{r'}(t'')\right)^{2}\right\rangle_{0}=\\
\left\langle\sum\limits_{a=1}^{2}
\left(\theta_{am}^{r}(t')\pm\theta_{am'}^{r'}(t'')\right)^{2}
+\delta_{ij}\sum\limits_{a=3}^{4}\left(\theta_{am}^{r}(t')\pm\theta_{am'}^{r'}(t'')\right)^{2}\right.\\
\left.+(1-\delta_{ij})\sum\limits_{a=3}^{4}\left(\phi_{am}^{r}(t')\pm\phi_{am'}^{r'}(t'')\right)^{2}\right\rangle_{0}.
\end{multline}
Note that the correlator above depends on the different processes
of interband $i\neq j$ or intraband $i=j$ scatterings which,
however, leads to the same correlation functions and only the
scattering phases hidden in $U^{\rm co}$ distinguish the different
processes. Also note that the correlation functions do not depend
on the spin direction $s$.   We find [e.g or the $\theta$-fields
(similar for $\phi$-fields)]
\begin{multline}
\label{generalc}
\left\langle\left(\theta_{am}^{r}(t)\pm\theta_{am'}^{r'}(0)\right)^{2}\right\rangle_{0}=\\
\pm
2C_{a|mm'}^{\theta\theta}(t)+2C_{a|mm}^{\theta\theta}(0)\\
\pm\frac{i}{2}(r'-r)\left[R_{a|mm'}^{\theta\theta}(t)-R_{a|m'm}^{\theta\theta}(-t)\right]\\
\pm\frac{i}{2}(r+r')\left[R_{a|mm'}^{\theta\theta}(t)+R_{a|m'm}^{\theta\theta}(-t)\right],
\end{multline}
where we have simplified the notation for clarity of the
presentation: $C_{a|mm'}^{\theta\theta}(t)\equiv
C_{a}^{\theta\theta}(x_{m},x_{m'};t)$ and
$R_{a|mm'}^{\theta\theta}(t)\equiv
R_{a}^{\theta\theta}(x_{m},x_{m'};t)$. We note that only the --
sign in Eq.~(\ref{generalc}) contributes as
$\exp[-(1/2)\langle(\theta_{am}^{r}(t)+\theta_{am'}^{r'}(0))^{2}\rangle]\propto
\exp(-\infty)$ due to the first line of the RHS in
Eq.~(\ref{generalc}). This is a direct consequence of particle
conservation \cite{particleconservation} since the + sign option
comes from terms like
$\langle\psi_{L}^{\dagger}\psi_{R}\psi_{L}^{\dagger}\psi_{R}\rangle_{0}$.
The general form of the noise described by  Eq.~(\ref{fullomega})
can be split into a sum of a clean limit with no
 backscattering corrections $S^{0}(x,\omega)$ and the
 backscattered correction $S_{I}(x,\omega)$ which we refer to as
 the impurity noise. The clean limit can be written in a more
 standard form \cite{Dolcini} using the relation between retarded and correlation
 function
\begin{equation}
\label{cleannoise}
S^{0}(x,\omega)=G_{0}\omega\coth\left(\frac{\beta\omega}{2}\right){\rm
Re}\,\sigma_{0}(x,x;\omega),
\end{equation}
where ${\rm Re}$ means real-part and we have introduced the
dimensionless conductivity \cite{conductivity} of the clean system
without the backscattering
$\sigma_{0}(x,y;\omega)=(2i\omega/\pi)R_{1}^{\theta\theta}(x,y;\omega)$.
The impurity contribution to the noise $S_{I}(x,\omega)$ can be
calculated using Eq.~(\ref{fullomega}) and Eqs.~(\ref{f1}) and
(\ref{generalc}). We obtain after some calculation the general
result valid for all temperatures, frequencies, gate voltages and
to leading order in the backscattering Hamiltonian $H_{\rm bs}$
(in units of $G_{0}\hbar/t_{\text{\textsc{f}}}$)
\begin{widetext}
\begin{multline}
\label{Vgnoise} S_{I}(x,\omega)=-2\coth\left(\frac{{\tilde
\omega}}{2\Xi}\right)\sum\limits_{m}U_{m}^{\rm in}\,{\rm
Re}\,\sigma_{0}(x,x_{m};\omega)\\
\times {\rm Im}\left[\sigma_{0}(x,x_{m};\omega)\,\int d\tau\,
e^{{\bf C}_{11}(\tau)}\sin[{\bf
R}_{11}(\tau)/2]\left(1-e^{i{\tilde \omega}
\tau}\right)\cos(v\tau)\right]\\
-\frac{1}{2}\sum\limits_{m}U_{m}^{\rm
in}\,|\sigma_{0}(x,x_{m};\omega)|^2\,\sum\limits_{r=\pm}\coth\left(\frac{v+r{\tilde
\omega}}{2\Xi}\right)\,\int d\tau\,e^{{\bf C}_{11}(\tau)}\sin[{\bf
R}_{11}(\tau)/2]\sin[(v+r{\tilde \omega})\tau]\nonumber\\
\end{multline}
\begin{multline}
-\coth\left(\frac{{\tilde \omega}}{2\Xi}\right)\sum\limits_{mm'}\,{\rm Re}\,\sigma_{0}(x,x_{m};\omega)\\
\times{\rm Im}\left\{\sigma_{0}(x,x_{m'};\omega)\,\int
d\tau\,e^{{\bf C}_{12}(\tau)}\,\sin[{\bf
R}_{12}(\tau)/2]\left(\delta_{mm'}-(1-\delta_{mm'})e^{i{\tilde
\omega} \tau}\right)
\bigl[U^{\rm co}\,\cos(v\tau)-V^{\rm co}(1-2\delta_{1m'})\sin(v\tau)\bigr]\right\}\\
+\frac{1}{2}\,{\rm
Re}\left\{\sigma_{0}(x,x_{1};\omega)^{*}\sigma_{0}(x,x_{2};\omega)\,\int
d\tau\,e^{{\bf C}_{12}(\tau)}\cos[{\bf R}_{12}(|\tau|)/2]
\,e^{i{\tilde \omega} \tau}\bigl[U^{\rm co}\cos(v\tau)-V^{\rm
co}\sin(v\tau)\bigl]\right\}.
\end{multline}
\end{widetext}
In Eq.~(\ref{Vgnoise}) we have used ${\bf
R}_{mm'}(\tau)=R_{mm'}^{\rm I}(\tau)+3 R_{mm'}^{\rm F}(\tau)$ and
a similar definition holds for ${\bf C}_{mm'}(\tau)$. The
superscripts ${\rm I}$ and ${\rm F}$ denote interacting and free,
respectively.  The interacting functions are
$\theta_{1}\theta_{1}$-correlations whereas the free functions
come from correlations of the non-interacting modes $a=2,3,4$. In
the main text we give the slightly more compact result for the
case where the coherent (FP)-contribution is maximum, i.e. when
$|U^{\rm co}|$ is maximum as a function of gate voltage. In
Eq.~(\ref{Vgnoise}) we have introduced $V^{\rm co}$ which is just
$U^{\rm co}$ with $\cos(V_{\rm g}L+2\Delta_{ij})$ replaced by
$\sin(V_{\rm g}L+2\Delta_{ij})$. We see that at finite frequency
$\omega$, the impurity noise becomes sensitive to the real and
imaginary parts of the conductance $\sigma_{0}(x,x_{m};\omega)$
which contains the multiple reflections of the charge mode $a=1$
at the inhomogeneity of $g$ where the SWNT is connected to the
non-interacting reservoirs. The complete noise as a function of
frequency and bias voltage is therefore a complicated
superposition of ordinary FP-oscillations described by the time
integrals which are influenced by both, voltage and frequency and
exhibit by all four modes whereas the additional frequency
response due to $\sigma_{0}(x,x_{m};\omega)$ is only sensitive to
the total charge mode $a=1$. For clarity, we give here the
explicit form of real and imaginary-parts of the retarded Green's
function connecting the two barriers with the measurement point
$x$ (assumed to be in the right lead). For the retarded function
with $x_{m}=-L/2$ we obtain from Eq.~(\ref{retardedfu})
\begin{widetext}
\begin{multline}
\label{p1}
R_{1}^{\theta\theta}(x,x_{1};\omega)=\frac{\pi}{2\omega}\frac{1-\gamma^2}{1-2\gamma^{2}\cos(2\omega
 t_{c})+\gamma^{4}}
\left\{\sin\left[\omega
t_{\text{\textsc{f}}}(g+d)\right]+\gamma^2\sin\left[\omega
t_{\text{\textsc{f}}}(g-d)\right]\right\}\\
-i\,\frac{\pi}{2\omega}\frac{1-\gamma^2}{1-2\gamma^{2}\cos(2\omega
 t_{c})+\gamma^{4}}
\left\{\cos\left[\omega
t_{\text{\textsc{f}}}(g+d)\right]-\gamma^2\cos\left[\omega
t_{\text{\textsc{f}}}(g-d)\right]\right\}.
\end{multline}
For the retarded function with $x_{m}=+L/2$ we obtain
\begin{multline}
\label{p2} R_{1}^{\theta\theta}(x,x_{2};\omega)=
\frac{\pi}{2\omega}\frac{1-\gamma}{1-2\gamma^{2}\cos(2\omega
 t_{c})+\gamma^{4}}
\left\{\sin\left(\omega
t_{\text{\textsc{f}}}d\right)\,[1+\gamma(1-\gamma)\cos(2\omega
t_{c})-\gamma^3]\right.
 \left.+\gamma(1+\gamma)\cos\left(\omega
t_{\text{\textsc{f}}}d\right)\,\sin(2\omega t_{c})\right\}\\
\\
-i\,\frac{\pi}{2\omega}\frac{1-\gamma}{1-2\gamma^{2}\cos(2\omega
 t_{c})+\gamma^{4}}\left\{\cos\left(\omega
t_{\text{\textsc{f}}}d\right)[1+\gamma(1-\gamma)\cos(2\omega
t_{c})-\gamma^3]\right. -\left.\gamma(1+\gamma)\sin\left(\omega
t_{\text{\textsc{f}}}d\right)\,\sin(2\omega t_{c})\right\}.
\end{multline}
\end{widetext}
 In Eqs.~(\ref{p1}) and (\ref{p2}) we have introduced the
distance from the measurement point $x$ to the nearest metal
contact-SWNT interface (here $x_{2}=L/2$) in units of the length
$L$ of the nanotube, i.e. $d=(x-x_{2})/L$.

For completeness, we also show a more direct way to obtain the
low-frequency noise Eq.~(\ref{zeronoise}) starting from
Eq.~(\ref{fullomega}) using a low-frequency expansion in $\omega$.
For this we consider the term
$\sum_{rr'}rr'f_{m}^{r}(x,\omega)[f_{m}^{r}(x,-\omega)-\exp(-i\omega
t)f_{m'}^{r'}(x,-\omega)]$ in the limit $\omega\rightarrow 0$. To
proceed in the evaluation we first make a straightforward
expansion of the above expression. We write the function
$f_{m}^{r}(x,\omega)$ in terms of real- and imaginary parts of the
retarded function $R_{1}^{\theta\theta}(x,x_{m};\omega)$ as
\begin{multline}
\label{fdecompose}
f_{m}^{r}(x,\omega)=\omega\left[\left(\frac{r}{2}-\coth\left(\frac{\beta\omega}{2}\right)\right)\,{\rm
Im}\,R_{1}^{\theta\theta}(x,x_{m};\omega)\right.\\
\left.\qquad+\frac{i}{2}r{\rm
Re}\,R_{1}^{\theta\theta}(x,x_{m};\omega)\right].
\end{multline}
 We now use the
low-frequency behavior of the real- and imaginary part of the
retarded functions in Eq.~(\ref{fdecompose}) which we express as
\begin{equation}
{\rm Re}\,R_{1}^{\theta\theta}(x,x_{m};\omega)=R_{1}^{\theta
0}(x,x_{m})+{\cal O}(\omega^2)
\end{equation}
and
\begin{equation}
{\rm
Im}\,R_{1}^{\theta\theta}(x,x_{m};\omega)=-\frac{\pi}{2\omega}+R_{1}^{\theta
1}(x,x_{m})\omega,
\end{equation}
where $R_{1}^{\theta 0}(x,x_{m})$ and $i\,R_{1}^{\theta
1}(x,x_{m})$ are the zeroth-order and 1st-order expansion
coefficient of $R_{1}^{\theta\theta}(x,x_{m};\omega)$,
respectively. They depend on $x$, $x_{m}$ and $g$ but we find that
these terms will not contribute to the zero-frequency limit of
noise. As a consequence, the low-frequency noise is independent on
the position of measurement. In this limit, we obtain for the
frequency dependent part of Eq.~({\ref{fullomega})

\begin{align}
\label{expansion}
&rr'\,f_{m}^{r}(x,\omega)\left[f_{m}^{r}(x,-\omega)-\exp(-i\omega
t)f_{m'}^{r'}(x,-\omega)\right]\nonumber\\
\nonumber\\
 =&(r-r')\left(\frac{\pi}{2}\right)^{2}k_{\rm
B}T\left(\frac{1}{\omega}-it\right)-it\,rr'\frac{(\pi
k_{\rm B}T)^{2}}{\omega}\nonumber\\
-&\frac{rr'}{2}(\pi
k_{\rm B}T)^{2}t^{2}-(1-rr')\left(\frac{\pi}{4}\right)^{2}\nonumber\\
+&2\pi\left(k_{\rm B}T\right)^2 rr'\left(R_{1}^{\theta
1}(x,x_{m})-R_{1}^{\theta 1}(x,x_{m'})\right)\nonumber\\
-&i\frac{\pi}{2}k_{\rm B}T\left(r'R_{1}^{\theta
0}(x,x_{m})-rR_{1}^{\theta 0}(x,x_{m'})\right)\nonumber\\
 +&{\cal O}(\omega).
\end{align}
By performing the sum over $rr'$ as well as the sum over $mm'$ we
find that the time integral in Eq.~(\ref{fullomega}) yields zero
for the terms associated with the $1/\omega$ contributions in
Eq.~(\ref{expansion}). Therefore, the limit $\omega\rightarrow 0$
is well defined. Note also that the voltage term
$\cos(eVt+\varphi_{mm'}^{{\bf
n}})=\cos(eVt)\cos(\varphi_{mm'}^{{\bf
n}})-\sin(eVt)\sin(\varphi_{mm'}^{{\bf n}})$ where, due to the
symmetry in the sum over $mm'$, only the cosine-terms contribute.
Using Eq.~(\ref{expansion}) in Eq.~(\ref{fullomega}) for the noise
we obtain the low-frequency noise presented in
Eq.~(\ref{zeronoise}).

\section{Derivation of retarded and correlation functions}
 In this appendix, we outline the derivation of the retarded
Green's functions which are calculated in the equilibrium system
and without the backscattering ($H_{\rm bs}$=0). We choose to
calculate the temperature Green's function first and then rotate
back to real time (Wick rotation) which gives us the retarded
function.

We start with deriving the action from the Hamiltonian
$H_{\text{\textsc{swnt}}}$
\begin{equation}
{\cal L}(\Pi_{a},\theta_{a})=\int dx
\sum\limits_{a}\,\left[\Pi_{a}(x){\dot\theta}_{a}(x)-H_{\text{\textsc{swnt}}}(\Pi_{a},\theta_{a})\right].
\end{equation}
The action is defined as the time integrated Lagrangian $S=\int dt
{\cal L}(t)$.  We then change to imaginary time $\tau=it$ and
introduce the Euclidean action by the standard identification
$-S_{E}=iS(it\rightarrow \tau)$. This immediately gives
\begin{align}
S_{E}=&\frac{1}{2}\sum\limits_{a}\int_{0}^{\beta}d\tau \int
dx\nonumber\\
\times&\left[i\frac{2}{\pi}\partial_{x}\phi_{a}\partial_{\tau}\theta_{a}+\frac{v_{\text{\textsc{f}}}}{\pi}[(\partial_{x}\phi_{a})^2+\frac{1}{g_{a}^2}(\partial_{x}\theta_{a})^2]\right].\nonumber\\
\end{align}
To calculate time-ordered correlation functions $\langle {\hat
T}{\hat A}({\bf x}){\hat B}({\bf x}')\rangle$ where ${\hat A}$ and
${\hat B}$ are any function of operators ${ \theta}_{a}$ and ${
\phi}_{a}$ we can use the functional integral approach
\begin{equation}
\label{functionalintegral}
 \langle {\hat T}{\hat A}({\bf x}){\hat B}({\bf
x}')\rangle=\frac{1}{Z}\prod_{a}\int{\cal
D}[\phi_{a}\theta_{a}]\,A({\bf x})B({\bf
x}')e^{-S_{E}(\theta_{a},\phi_{a})},
\end{equation}
where ${\bf x}=(x,\tau)$ and $Z=\prod_{a}\int{\cal
D}[\phi_{a}\theta_{a}]\,e^{-S_{E}(\theta_{a},\phi_{a})}$ is the
partition function. Here, since the bosonic fields are hermitesch,
the functional integral is over real-valued fields. If the
operators ${\hat A}$  and ${\hat B}$ are only functions of one of
the field-type, i.e. only a function of either  ${\theta_{a}}$ or
${\phi_{a}}$, we can integrate out the other variable to get an
effective action which only depends on one of the variables. To do
this we use the result for Gaussian integration over
multidimensional real variables $x_{i},\,i=1,...,N$
\begin{equation}
\label{Gaussian} \prod_{i}\int
dx_{i}\,e^{-\frac{1}{2}\sum\limits_{ij}\,x_{i}A_{ij}x_{j}+\sum\limits_{i}\lambda_{i}x_{i}}=\frac{(2\pi)^{N/2}}{\sqrt{{\rm
det
A}}}\,e^{\frac{1}{2}\sum\limits_{ij}\,\lambda_{i}A_{ij}^{-1}\lambda_{j}},
\end{equation}
where det means the determinant of the real symmetric and positive
definite matrix $A_{ij}$. We first derive the action for  the
$\theta_{a}-$ fields. Using partial integration in the action
$S_{E}$ we can bring the functional integral
Eq.~(\ref{functionalintegral}) to the form of Eq.~(\ref{Gaussian})
with identifying $\lambda_{i}\equiv
(i/\pi)\partial_{x}\partial_{\tau}\theta_{a}$ and the matrix
elements $A_{ij}\equiv
-(v_{\text{\textsc{f}}}/\pi)\partial_{x}^{2}\delta(x-x')\delta(\tau-\tau')$.
Note that the determinant will cancel with the similar integration
procedures in the partition function. Therefore, for calculating
correlation functions the explicit calculation of the determinant
in Eq.~(\ref{Gaussian}) is not needed. We then obtain the
effective action for the $\theta_{a}-$ fields
\begin{equation}
S^{\theta}_{E}=\frac{1}{2\pi}\sum_{a}\int_{0}^{\beta} d\tau\int
dx\left[\frac{v_{a}}{g_{a}}(\partial_{x}\theta_{a})^{2}+\frac{1}{g_{a}v_{a}}(\partial_{\tau}\theta_{a})^{2}\right].
\end{equation}
Similarly, for the effective action of $\phi_{a}-$fields we obtain
\begin{equation}
S^{\phi}_{E}=\frac{1}{2\pi}\sum_{a}\int_{0}^{\beta} d\tau\int
dx\left[v_{a}g_{a}(\partial_{x}\phi_{a})^{2}+\frac{g_{a}}{v_{a}}(\partial_{\tau}\phi_{a})^{2}\right].
\end{equation}
We can now use the effective actions to calculate the correlation
functions $\langle {\hat T}\theta_{a}({\bf x})\theta_{a}({\bf
x}')\rangle$ and $\langle {\hat T}\phi_{a}({\bf x})\phi_{a}({\bf
x}')\rangle$, respectively.\\
The relation between functional integrals and time-ordered
correlation functions Eq.~(\ref{functionalintegral})  (e.g. for
$\theta_{a}$-fields)
\begin{equation}
\label{functionalintegral2}
 \langle  {\hat T}{
\theta}_{a}({\bf x}){\theta}_{a}({\bf
x}')\rangle=\frac{1}{Z^{\theta_{a}}}\int{\cal
D}\theta_{a}\,\theta_{a}({\bf x})\theta_{a}({\bf
x}')e^{-S_{E}(\theta_{a})}
\end{equation}
can also be written as
\begin{equation}
 \langle {\hat  T}{
\theta}_{a}({\bf x}){ \theta}_{a}({\bf
x}')\rangle=\left.\frac{\delta^2}{\delta\lambda(\bf
x)\delta\lambda(\bf
x')}\,Z^{\lambda}_{\theta_{a}}\right|_{\lambda=0},
\end{equation}
where
\begin{equation}
Z_{\theta_{a}}^{\lambda}=\frac{1}{Z^{\theta_{a}}}\int{\cal
D}\theta_{a}\,e^{-S_{E}^{\theta_{a}}+\int d{\bf x}\,\lambda({\bf
x})\,\theta_{a}({\bf x})}
\end{equation}
is the generating functional for the $\theta_{a}$-field. Writing
the action $S_{E}^{\theta_{a}}$ as a bilinear form
$S_{E}^{\theta_{a}}=(1/2)\int d{\bf x}\int d{\bf
x'}\theta_{a}({\bf x})[{\hat G}_{a}^{\theta\theta}]^{-1}({\bf
x},{\bf x'})\theta_{a}({\bf x}')$ and using the result for
Gaussian integration Eq.~(\ref{Gaussian}) we conclude that
$\langle {\hat T}{\theta}_{a}({\bf x}){ \theta}_{a}({\bf
x}')\rangle={\hat G}_{a}^{\theta\theta}({\bf x},{\bf x'})$
accompanied with the operator statement
\begin{equation}
\left[{\hat G}_{a}^{\theta\theta}\right]^{-1}{\hat
G}_{a}^{\theta\theta}={\hat 1}.
\end{equation}
Using that the inverse Green's function operator ${\hat G}^{-1}$
is local in (imaginary)time and space, i.e. $\langle {\bf x}|{\hat
G}^{-1}|{\bf x}'\rangle={\hat D}({\bf x}) \delta({\bf x}-{\bf
x}')$,  leads us to the differential equation for the Green's
function in imaginary time
\begin{equation}
\label{operatoreq}
 {\hat
D}(x,\tau)G(x,\tau;x',\tau')=\delta(x-x')\delta(\tau-\tau').
\end{equation}
Eq.~(\ref{operatoreq}) clearly shows that the Green's function is
symmetric in ${\bf x}$ and ${\bf x}'$. Explicitly, we obtain the
differential-operators ${\hat D}^{\theta\theta}_{a}({\bf
x})=-\frac{1}{\pi}[\frac{1}{v_{a}g_{a}}\partial_{\tau}^{2}+\partial_{x}\frac{v_{a}}{g_{a}}\partial_{x}]$
and ${\hat D}^{\phi\phi}_{a}({\bf
x})=-\frac{1}{\pi}[\frac{g_{a}}{v_{a}}\partial_{\tau}^{2}+\partial_{x}v_{a}g_{a}\partial_{x}]$.
Due to  the inhomogeneity of the charge mode $a=1$, its Green's
functions are not translational invariant. The time translation
invariance although still holds, i.e.
$G(x,\tau;x',\tau')=G(x,x';\tau-\tau')$  and we can transform to
frequency space using the Fourier expansion for boson Matsubara
Green's functions
\begin{equation}
G(x,x';\tau)=\frac{1}{\beta}\sum\limits_{\omega_{n}}e^{-i\omega_{n}\tau}G(x,x';\omega_{n}),
\end{equation}
where the sum is over the Matsubara frequencies $\omega_{n}=2\pi
n/\beta$ with $n=0,\pm 1,\pm 2$.  We then obtain a differential
equation for the Fourier component of the Matsubara Green's
function
\begin{equation}
\label{Greentheta}
\left(\frac{1}{g_{a}v_{a}}\omega_{n}^{2}-\partial_{x}\frac{v_{a}}{g_{a}}\partial_{x}\right)G_{a}^{\theta\theta}(x,x';\omega_{n})=\pi\delta(x-x'),
\end{equation}
and
\begin{equation}
\label{Greenphi}
\left(\frac{g_{a}}{v_{a}}\omega_{n}^{2}-\partial_{x}v_{a}g_{a}\partial_{x}\right)G_{a}^{\phi\phi}(x,x';\omega_{n})=\pi\delta(x-x').
\end{equation}
The solution to these partial differential equations for the case
$|x'|< L/2$ can be found by the ansatz \cite{MaslovStone}(for the
$\theta\theta$-contribution)
\begin{eqnarray}
\label{Greensolution}
&&G_{a}^{\theta\theta}(x,x';\omega_{n})=\nonumber\\
&&\nonumber\\ &&\left\{\begin{array}{lll}
A e^{|\omega_{n}|x/v_{\text{\textsc{f}}}};& x\leq -L/2\\
B e^{|\omega_{n}|x/v_{a}}+C e^{-|\omega_{n}|x/v_{a}};&-L/2 <x\leq
x'<L/2\\
D e^{|\omega_{n}|x/v_{a}}+E e^{-|\omega_{n}|x/v_{a}};&-L/2 <
x'<x\leq L/2\\
F e^{-|\omega_{n}|x/v_{\text{\textsc{f}}}};& x>L/2\\
\end{array}
\right.\nonumber\\
\end{eqnarray}
These solutions satisfy the boundary condition
$G_{a}^{\theta\theta}(\pm \infty,x';\omega_{n})=0$. The
coefficients $A-F$ are functions of $x'$ and $\omega_{n}$ and can
be found from the following boundary conditions
\begin{displaymath}
\begin{array}{lll}
{\bullet} \,G_{a}^{\theta\theta}(x,x';\omega_{n})\, \,{\rm is\,\,
continuous\,\,
everywhere}.\\
{\bullet}\,\frac{v_{a}(x)}{g_{a}(x)}\partial_{x}G_{a}^{\theta\theta}(x,x';\omega_{n})\,{\rm
\,\,is \,\,continuous \,\,at }\,\,x=L/2,-L/2. \\
{\bullet}\,-\frac{v_{a}(x)}{g_{a}(x)}\partial_{x}G_{a}^{\theta\theta}(x,x';\omega_{n})\,\,{\rm
has\,\,a\,\,step\,\,of\,\,height}\,\,\pi\,\,{\rm at}\,\,x=x'.
\end{array}
\end{displaymath}
These three conditions lead to the following set of equations
which can be used to determine all constants $A-F$:
\begin{displaymath}
\begin{array}{l}
A=B e^{|\omega_{n}|L(1-g_{a})/2v_{\text{\textsc{f}}}}+C e^{|\omega_{n}|L(1+g_{a})/2v_{\text{\textsc{f}}}}\\
F=D e^{|\omega_{n}|L(1+g_{a})/2v_{\text{\textsc{f}}}}+E
e^{|\omega_{n}|L(1-g_{a})/2v_{\text{\textsc{f}}}}\\
C+B e^{2|\omega_{n}|x'/v_{a}}=E+D e^{2|\omega_{n}|x'/v_{a}}\\
g_{a} A=B e^{|\omega_{n}|L(1-g_{a})/2v_{\text{\textsc{f}}}}-C
e^{|\omega_{n}|L(1+g_{a})/2v_{\text{\textsc{f}}}}\\
g_{a}F=Ee^{|\omega_{n}|L(1-g_{a})/2v_{\text{\textsc{f}}}}-D
e^{|\omega_{n}|L(1+g_{a})/2v_{\text{\textsc{f}}}}\\
B-D+(E-C)e^{-2|\omega_{n}|x'/v_{a}}=\frac{g_{a}\pi}{|\omega_{n}|}e^{-|\omega_{n}|x'/v_{a}}\\
\end{array}
\end{displaymath}

The retarded Green's function
$R^{\theta\theta}_{a}(x,t;x',t')=-i\Theta(t-t')\langle[\theta_{a}(x,t),\theta_{a}(x',t')]\rangle$
is obtained from the Matsubara Green's function via the analytic
continuation
\begin{equation}
R^{\theta\theta}_{a}(x,x';\omega)=-\,G_{a}^{\theta\theta}(x,x';\omega_{n})\bigr|_{i\omega_{n}\rightarrow
\omega+i\delta},
\end{equation}

with $\delta=0^{+}$. The analytic continuation is performed from
the positiv imaginary axis where the function is
$G_{a}^{\theta\theta}(x,x';\omega_{n})$ with $\omega_{n}>0$ to
just above the real axis where it equals
$R^{\theta\theta}_{a}(x,x';\omega)$. This amounts to the
replacement $|\omega_{n}|\rightarrow -i \omega+\delta$ in
$G_{a}^{\theta\theta}(x,x';\omega_{n})$ to obtain the retarded
function $R^{\theta\theta}_{a}(x,x';\omega)$. The function for the
$\phi_{a}$-fields can be  obtained from the solution for the
$\theta_{a}$-fields by the substitution $g_{a}\rightarrow 1/g_{a}$
which is evident from  the differential equations Eqs.
(\ref{Greentheta}) and (\ref{Greenphi}). For the retarded
functions $R^{\theta\theta}_{a|12}(\omega)$ and
$R^{\theta\theta}_{a|11}(\omega)$ we need the solutions in
Eq.~(\ref{Greensolution}) with $-L/2\leq x \leq L/2$. Since the
Green's function is continuous everywhere we get also the correct
solution at the boundary where $x'=\pm L/2$. We obtain for general
$|x,x'|\leq L/2$ and for the interacting mode $a=1$
\begin{multline}
\label{retarded1}
 R^{\theta\theta}_{1}(x,x';\omega)=\frac{-i\pi
g}{2{\bar \omega}}\biggl\{e^{i{\bar\omega}t_{c}|x-x'|/L}
+\frac{\gamma}{e^{-i2{\bar \omega}t_{c}}-\gamma^2}\\
\times\sum\limits_{r=\pm}\left[e^{-i{\bar
\omega}t_{c}[1-r(x+x')/L]}+\gamma e^{i{\bar
\omega}t_{c}r(x-x')/L}\right]\biggr\},
\end{multline}
where ${\bar \omega}=\omega+i\delta$, and
$t_{c}=Lg/v_{\text{\textsc{f}}}$. We further introduced
$\gamma=(1-g)/(1+g)$ which can be interpreted as the reflection
coefficient for an incoming current flux traversing the
reservoir-nanotube interface \cite{SafiSchulz}(i.e. the
inhomogeneity of $g$). We also need the retarded functions in real
time which we get by Fourier transforming Eq.~(\ref{retarded1}).
Using $(1-\gamma^2 e^{i2\omega
t_{c}})^{-1}=\sum_{k=0}^{\infty}\gamma^{2k}\,e^{i 2k\omega t_{c}}$
and a high-energy cut-off function $e^{-|\omega|/\omega_{0}}$
 we obtain for $x=x'=\pm L/2$
\begin{multline}
r_{1|11}^{\theta\theta}(t)=-\frac{\pi}{2}(1-\gamma)\\
\times\left\{\Theta_{\omega_{0}}(t)+\frac{1+\gamma}{\gamma}\sum\limits_{k=1}^{\infty}\gamma^{2k}\Theta_{\omega_{0}}(t-2kt_{c})\right\},
\end{multline}
and for the cross-terms $x=L/2(-L/2)$, $x'=-L/2(L/2)$ describing
the FP-interference we obtain
\begin{equation}
r_{1|12}^{\theta\theta}(t)=-\frac{\pi}{2}(1-\gamma^2)\sum\limits_{k=0}^{\infty}\gamma^{2k}\Theta_{\omega_{0}}[t-(2k+1)t_{c}].
\end{equation}
The smeared step-function is defined as
$\Theta_{\omega_{0}}(t)=(1/\pi)\arctan(\omega_{0}t)+1/2$. If we
keep the cut-off finite the correct retarded function is obtained
by the combination
$R_{1|mm'}^{\theta\theta}(t)=\theta(t)[r_{1|mm'}^{\theta\theta}(t)-r_{1|m'm}^{\theta\theta}(-t)]$.
Note that the retarded Green's functions are temperature
independent. The temperature dependence is completely contained in
the correlation functions to be derived next. First note that we
are dealing with equilibrium properties, and therefore the
correlation function is connected to the retarded function via the
fluctuation-dissipation theorem
\begin{equation}
C_{a|mm'}^{\theta\theta}(\omega)=\frac{i}{2}\coth(\beta\omega/2)[R_{a|mm'}^{\theta\theta}(\omega)-R_{a|m'm}^{\theta\theta}(-\omega)].
\end{equation}
We give here the results for the $\theta\theta$-correlations. The
corresponding results for the $\phi\phi$-correlations are obtained
by the replacement: $g_{a}\rightarrow 1/g_{a}$. We split the
temperature dependence in a $T=0$ part plus the temperature
corrections, $C_{a|mm'}^{\theta\theta}(t)=C_{a|mm'}^{\theta\theta
0}(t)+C_{a|mm'}^{\theta\theta {\rm T}}(t)$. Note that we can
decompose
\begin{equation}
\coth\left(\frac{\beta\omega}{2}\right)=1+\frac{2e^{-\beta\omega}}{1-e^{-\beta\omega}}.
\end{equation}
For positive frequencies $\omega$  we can write
$1/(1-e^{-\beta\omega})$ as a geometric series valid for all
temperatures. We then obtain
$\coth(\beta\omega/2)=1+2\sum_{n=0}^{\infty}e^{-\beta\omega(n+1)}$
which corresponds to the two terms contributing either at zero
temperature  $\coth(\beta\omega/2)=1$ or to the finite temperature
corrections $
\coth(\beta\omega/2)=2\sum_{n=0}^{\infty}e^{-\beta\omega(n+1)}$.
At zero temperature we then obtain
\begin{multline}
\label{A2} C_{a|12}^{\theta\theta
0}(t)=-\frac{1}{8}(1-\gamma^2)\\
\times\sum\limits_{k=0}^{\infty}\gamma^{2k}\sum\limits_{r=\pm}\ln\left[\omega_{0}^{-2}+(t+r(2k+1)t_{c})^{2}\right],
\end{multline}
and
\begin{multline}
\label{A3} C_{a|11}^{\theta\theta
0}(t)=-\frac{1-\gamma}{4}\biggl\{\ln(\omega_{0}^{-2}+t^2)\\
+\frac{1+\gamma}{2\gamma} \sum\limits_{k=1}^{\infty}\gamma^{2k}
\sum\limits_{r=\pm}\ln\left[\omega_{0}^{-2}+(t+r2kt_{c})^{2}\right]\biggr\}.
\end{multline}
Note in Eqs. (\ref{A2}) and (\ref{A3}) we omitted a time and
space-independent constant which does not contribute to the
relevant combination  $C_{a|mm'}^{\theta\theta
}(t)-C_{a|mm}^{\theta\theta }(0)$. For the finite temperature
correction $C_{a|12}^{\theta\theta {\rm T}}$ we obtain
\begin{multline}
C_{a|12}^{\theta\theta {\rm T}}(t)=\frac{1-\gamma^{2}}{4}\\
\times\sum\limits_{k=0}^{\infty}\gamma^{2k}\sum\limits_{r=\pm}\ln\left|\frac{\Gamma\left(\frac{1}{\omega_{0}\beta}+1+i\frac{t+rt_{c}(2k+1)}{\beta}\right)}{\Gamma\left(1+\frac{1}{\omega_{0}\beta}\right)}\right|^{2}.
\end{multline}
In the finite temperature correlation functions it is allowed to
perform the limit $\omega_{0}\rightarrow \infty$ as the finite
temperature plays the role of the cut-off. This is true as long as
$k_{\rm B}T$ is small compared to the high-energy cut-off
$\epsilon_{0}$. After doing so, we can use that
$|\Gamma(1+ix)|^{2}=\pi x/\sinh(\pi x)$ for $x$ real to obtain the
simpler form
\begin{multline}
C_{a|12}^{\theta\theta {\rm T}}(t)=\frac{1-\gamma^{2}}{4}\\
\times\sum\limits_{k=0}^{\infty}\gamma^{2k}\sum\limits_{r=\pm}\ln\left[\frac{\pi(t+r(2k+1)t_{c})}{\beta\sinh[\pi(t+r(2k+1)t_{c})/\beta]}\right].
\end{multline}
For the autocorrelation functions $(m=m')$ we obtain
\begin{multline}
C_{a|mm}^{\theta\theta {\rm
T}}(t)=\frac{1-\gamma}{2}\ln\left[\frac{\pi t}{\beta\sinh(\pi
t/\beta)}\right]\\
+\frac{1-\gamma^{2}}{4\gamma}\sum\limits_{k=1}^{\infty}\gamma^{2k}\sum\limits_{r=\pm}\ln\left[\frac{\pi(t+r2kt_{c})}{\beta\sinh[\pi(t+r2kt_{c})/\beta]}\right].
\end{multline}
The frequency representation of the retarded function
$R_{1}^{\theta\theta}(x,x_{m};\omega)$ given in
Eq.~(\ref{retardedfu}) where the measurement point $x\geq L/2$ is
chosen to be in the right lead and $x_{m}=\pm L/2$ can also be
obtained from Eq.~(\ref{Greensolution}) in the regime $x>L/2$.



\begin{references}
\bibitem{Kleinfactor}
In order for the Fermion operators  Eq.~(\ref{bd}) to explicitly
anticommute one has to introduce Kleinfactors, which, however, are
not crucial in this work.
\bibitem{conductivity}
The dimensionfull conductivity is $G_{0}\sigma_{0}(x,x';\omega)$.
The reason to introduce a dimensionless quantity is that we prefer
to plot all results in dimensionless quantities.
\bibitem{particleconservation}
In other words, since all total chiral particle number operators
commute with $H_{\text{\textsc{swnt}}}$ they are all constants of
motion and the eigenstates can be chosen to be simultaneous
eigenstates of all chiral total particle number operators.
\end{references}
\end{document}